\documentclass[12pt,preprint]{aastex}
\pdfoutput=1
\usepackage{epsfig}
\usepackage{graphicx}
\usepackage{amsmath}
\usepackage{natbib}
\usepackage{color}
\usepackage{./emulateapj5}
\usepackage{./onecolfloat}
\usepackage{./apjfonts}
\usepackage{./subfigure}

\usepackage{enumerate}
\newif\ifAMStwofonts
\AMStwofontstrue

\def\tilde{~}

\def\rvir{R_{\rm vir}}

\def\rfive{R_{\rm 500}}

\def\rtwofive{R_{\rm 2500}}

\def\rtwo{R_{\rm 200}}
\def\mtrue{M_{\rm true}}
\def\mhe{M_{\rm HE}}

\def\msun{\rm{\,M_{\odot}}}

\def\msunh{\,h^{-1}\rm{\,M_{\odot}}}

\def\kev{\rm{\,keV}}

\def\kpc{\rm{\,kpc}}
\def\kpch{\,h^{-1}\rm{\,kpc}}
\def\mpc{\rm{\,Mpc}}

\def\gacc{${\cal G}_r$}
\def\hacc{${\cal H}_r$}
\def\ratio{{\cal G}_r/{\cal H}_r}
\def\dHE{\delta_{\rm HE}}

\def\tmw{T_{\rm mw}}
\def\tsl{T_{\rm sl}}

\newcommand{\be}{\begin{equation}}
\newcommand{\ee}{\end{equation}}
\def\aap{A\&A}

\def\apjl{ApJ}

\def\apjs{ApJS}

\shorttitle{HE in simulated clusters}
\shortauthors{V. Biffi et al.}
\begin{document}
\twocolumn[%
\title{On the nature of hydrostatic equilibrium in
galaxy clusters}

\author{
V.~Biffi\altaffilmark{1,2},
S.~Borgani\altaffilmark{1,2,3},
G.~Murante\altaffilmark{2},
E.~Rasia\altaffilmark{2,4},
S.~Planelles\altaffilmark{1,2,5},
G.L.~Granato\altaffilmark{2},
C.~Ragone-Figueroa\altaffilmark{6},
A.M.~Beck\altaffilmark{7},
M.~Gaspari\altaffilmark{8},
K.~Dolag\altaffilmark{7}
}
\affil{$^{1}$ Astronomy Unit, Department of Physics, University of Trieste, via Tiepolo 11, I-34131 Trieste, Italy; biffi@oats.inaf.it}

\affil{$^{2}$ INAF, Osservatorio Astronomico di Trieste --- OATs, via Tiepolo 11, I-34131 Trieste, Italy}

\affil{$^3$ INFN --- National Institute for Nuclear Physics, Via Valerio 2, I-34127 Trieste, Italy}

\affil{$^4$ Department of Physics, University of Michigan, 450 Church St., Ann Arbor, MI 48109, USA}

\affil{$^5$ Departamento de Astronom{\'i}a y Astrof{\'i}sica, Universidad de Valencia, c/ Dr. Moliner, 50, 46100 - Burjassot (Valencia), Spain}

\affil{$^6$ Instituto de Astronom\'ia Te\'orica y Experimental (IATE), Consejo Nacional de Investigaciones Cient\'ificas y T\'ecnicas de la Rep\'ublica Argentina (CONICET), Observatorio Astron\'omico, Universidad Nacional de C\'ordoba, Laprida 854, X5000BGR, C\'ordoba, Argentina}

\affil{$^7$ University Observatory Munich, Scheinerstr. 1, D-81679 Munich, Germany}

\affil{$^8$ Department of Astrophysical Sciences, Princeton University, Princeton, NJ 08544, USA; Einstein and Spitzer Fellow}

\begin{abstract}
In this paper we investigate the level of hydrostatic equilibrium (HE)
in the intra-cluster medium of simulated galaxy clusters, extracted
from state-of-the-art cosmological hydrodynamical simulations
performed with the Smoothed-Particle-Hydrodynamic code GADGET-3.
These simulations include several physical processes, among which
stellar and AGN feedback, and have been performed with an improved
version of the code that allows for a better description of
hydrodynamical instabilities and gas mixing processes.  Evaluating the
radial balance between the gravitational and hydrodynamical forces,
via the gas accelerations generated, we effectively examine the level
of HE in every object of the sample, its dependence on the radial
distance from the center and on the classification of the cluster in
terms of either cool-coreness or dynamical state. We find an average
deviation of 10--20\% out to the virial radius, with no evident
distinction between cool-core and non-cool-core clusters. Instead, we
observe a clear separation between regular and disturbed systems, with
a more significant deviation from HE for the disturbed objects.  The
investigation of the bias between the hydrostatic estimate and the
total gravitating mass indicates that, on average, this traces very
well the deviation from HE, even though individual cases show a more
complex picture. Typically, in the radial ranges where mass bias and
deviation from HE are substantially different, the gas is
characterized by a significant amount of random motions ($\gtrsim 30$
per cent), relative to thermal ones. As a general result, the HE-deviation and mass bias, at given interesting distance from the cluster center, are
not very sensitive to the temperature inhomogeneities in the gas.
\end{abstract}
]

%
%
\section{Introduction}\label{sec:intro}

As fair samples of the Universe, galaxy clusters are dominated in mass,  $\sim 80$~per cent,
by dark matter (DM) but also comprise a
significant amount of baryonic visible matter, in the form of galaxies and hot plasma
($\sim 5$ and $\sim 15$~per cent in mass, respectively).
In the accepted scenario of hierarchical structure formation, clusters
grow via smooth accretion processes as well as through merger events
\cite[see][for a review]{kravtsov2012}.
According to this theoretical framework, the hot intra-cluster medium
(ICM) is assumed to collapse within the cluster DM halo, get shock
heated during the assembly process, and finally settle with
temperatures  of order $10^7$--$10^8$\,K, reflecting the
depth of the potential well ($\sim 10^{14}$--$10^{15}\msun$).  The
dynamics of the intra-cluster gas can be described by the Euler
equation:
\begin{equation}\label{eq:euler}
\frac{d{\bf v}}{dt} = -\nabla \Phi -\frac{1}{\rho}\nabla P \,.
\end{equation}
Here, $P$ is the total gas pressure, $\Phi$ is the cluster potential and
\be\label{eq:lagr-deriv}
\frac{d{\bf v}}{dt} = \frac{\partial {\bf v}}{\partial t}
+ \left( {\bf v} \cdot \nabla \right){\bf v}
\ee
is the Lagrangian derivative of the velocity or the sum of the
acceleration and the inertia terms, respectively the first and second
term on the l.h.s.\ of Eq.~\ref{eq:lagr-deriv}.  The condition of
hydrostatic equilibrium (HE) is represented by
\be\label{eq:he0}
\frac{d{\bf v}}{dt} = 0\,.
\ee
\sloppy{This assumption implies that the net Lagrangian
  three-dimensional acceleration of the gas, resulting from the sum of
  hydrodynamical and gravitational forces, is null.  With our
  numerical study, we intend to investigate whether the condition
  expressed by Eq.~\eqref{eq:he0},
and so the balance between hydrodynamical and gravitational forces is
reliable in cosmological simulations of galaxy clusters,
when evaluated at typical, interesting distances from the cluster center.
In fact, the
  assumption of HE is key ingredient behind one of the most
  diffuse methods employed to measure the galaxy cluster mass, which
  is the crucial property to characterize a cluster for astrophysical
  as well as cosmological purposes.

Specifically, the reconstruction of the so-called hydrostatic mass
from X-ray observations of the ICM thermal properties
can be derived from Eq.~\eqref{eq:he0} re-formulated as
$$0 = - \nabla \Phi
-\frac{1}{\rho}\nabla P\,,$$
along with the additional assumptions of spherical symmetry and of a
purely thermal pressure support of the gas ($P=P_{\rm th}$). Under
these conditions, and further assuming the equation of state of an ideal
gas to hold for the ICM, one can derive the total mass from the
profiles of gas density~($\rho$) and temperature~($T$):
\be\label{eq:mhe} \mhe (<r) = -\frac{k_B T(r) r}{\mu G m_p} \left[
  \frac{d\log\rho(r)}{d\log r} + \frac{d \log T(r)}{d\log r} \right],
\ee
where $k_B$ is the Boltzmann constant, $\mu$ the mean molecular
weight, $G$ is the gravitational constant, and $m_p$ the proton mass.

For regular virialized galaxy clusters the above assumptions
are a reasonable representation of the gas state.
However, if any of the hypotheses done are not satisfied,
the hydrostatic mass might provide a biased estimate of the true
gravitating mass.

Observationally, the particular composition of galaxy clusters allows
us to observe them in many different wavelengths other than X-rays,
such as optical or millimetric bands, providing independent methods to
reconstruct their intrinsic structure and total mass~\cite[see
  e.g.][]{mahdavi2008,mahdavi2013,vonderlinden2014,sereno2015,applegate2015,simet2015,smith2016}.
Some of these approaches, such as the one based on optical
observations of the weak lensing effect, are less sensitive to the
complex non-gravitational processes that characterise the gas and have
been commonly used for comparisons to X-ray mass
estimates~\cite[e.g.][]{donahue2014}.
A mismatch between optical and X-ray mass measurements has been often
interpreted as lack of HE.
Nonetheless, it is important to notice that
the violation of {\it any} of the assumptions behind
Eq.~\eqref{eq:mhe} can lead to a bias in the mass estimate, even in the presence
of a perfect balance between gravity and pressure.

To this end, numerical studies based on
state-of-the-art cosmological hydrodynamical simulations of galaxy
clusters offer an optimal way of tackling the problem.
Several groups have explored the hydrodynamical stability of simulated
clusters, computing the thermal and non-thermal components derived
from Eqs.~\eqref{eq:euler} and~\eqref{eq:lagr-deriv}
that  contribute to the total support
against the cluster gravitational potential and, if neglected, induce
the hydrostatic mass bias.  In fact, hydrodynamical simulations show
that a non-negligible fraction of the ICM pressure support is due to
turbulent and bulk gas motions, and this should be taken into account for
the mass estimation based on hydrostatic
equilibrium~\cite[][]{rasia2004,lau2009,fang2009,vazza2009,biffi2011,suto2013,lau2013,gaspari2013,gaspari2014,nelson2014a}.
Previous attempts to specifically identify the principal sources of
bias led however to different conclusions, mainly because of
differences in the terminology, computational method or interpretation
of the mass terms involved~\cite[][]{suto2013,lau2013,shi2015}.
In~\cite{fang2009} the major source of additional pressure support
against gravity has been ascribed to gas rotational patterns,
especially in the center of relaxed systems, whilst in~\cite{lau2009}
the authors found a significantly higher contribution to the total
pressure support coming from random motions. More recently, numerical
investigations by~\cite{suto2013} and~\cite{nelson2014a} additionally
explored the possibility of a non-steady state of the gas,
i.e. $\partial {\bf v}/\partial t \neq 0$, in Eq.~\eqref{eq:lagr-deriv},
and assessed the importance of accounting for gas acceleration as
well.
The common finding of numerical works is that, even for very regular
clusters, the hydrostatic mass overall underestimates the true
gravitating mass by a typical factor of $10$--$20$ per cent~\cite[see
  e.g.][]{rasia2004,rasia2006,jeltema2008,piffaretti2008,meneghetti2010,nelson2012}.
Independently of this, also the presence of gas temperature
inhomogeneities can cause an additional bias in the temperature
estimate from current X-ray telescopes,
thus originating a total difference between X-ray derived and true
masses of up to $30$ per cent~\cite[e.g.,][]{rasia2014}.

Even if challenging, the detection of gas turbulent and bulk motions
will substantially improve with observations from next-generation
X-ray calorimeters, on board satellites such as
ASTRO-H\footnote{http://astro-h.isas.jaxa.jp/en/.}  or
Athena\footnote{http://www.the-athena-x-ray-observatory.eu/.}.  Their
unprecedented level of energy resolution will eventually allow us to
put tighter constraints on the ICM motions, measuring gas velocities
from the broadening and center shifts of heavy-ions emission lines in
the X-ray spectra down to few hundreds
km/s~\cite[][]{biffi2013,athena,athenaettori2013}.

\looseness=-1 Nonetheless, it is very difficult to measure corrections for the mass
bias and generally deviations from HE of the gas on a single
cluster base at intermediate-high redshift, where the spatial precision
is more limited, and a statistical approach is therefore
preferable.  In fact, a thorough investigation of the origin of HE
violation for individual cases can only be pursued via numerical
simulations, which grant access to the full three-dimensional thermal
and velocity structure of clusters and to their dynamical
history. Complementary to this, simulations can also be exploited to
provide general predictions for cluster populations selected on the
base of common thermal or dynamical properties, more similarly to the
observational approach.
Even though gas motions might be constrained in the next future, it
remains extremely challenging to observationally distinguish among the
intrinsic deviation from the assumption of the steady state and other
sources of mass bias (e.g., sphericity and non-thermal pressure).
Certainly, the general relation between the common definition of
hydrostatic mass bias and deviation from HE by means of numerical
studies is worth of further investigation.

Here, instead of focusing on the various terms that originate the
  mass bias, we rather aim at taking a step back with respect to the
  previous numerical studies and explore, from a more elementary
  perspective, the primary assumption of the hydrostatic equilibrium
  expressed by Eq.~\eqref{eq:he0} and intended as balance of
  gravitational and hydrodynamical force.  Despite the fact that the mass
  bias is the observable quantity, our numerical approach represents a
  unique chance to quantify the intrinsic deviation from hydrostatic
  equilibrium, and consequently its connection to the mass bias, its dependence on
  the cluster thermo-dynamical properties and its relation to the
  level of random and bulk motions in the gas.
Specifically, we propose to investigate the level of hydrostatic
equilibrium of the ICM in simulated clusters, expressed by the
condition~\eqref{eq:he0}, by exploring the three-dimensional gas
acceleration~field.

\sloppy{The use of state-of-the-art cosmological hydrodynamical simulations of
galaxy clusters allow us to {\it directly} evaluate $(i)$ the balance
between hydrodynamical and gravitational forces through the comparison
of the accelerations derived from the two terms, $(ii)$ the dependence
on the distance from the cluster center, and $(iii)$ the possible
connections to the thermo-dynamical state of the system.}

\looseness=-1
This paper is  organized as follows: we present the simulations of
galaxy clusters used for the present study in Section~\ref{sec:sims},
while in Section~\ref{sec:method} describe the method to evaluate the level of HE in the
simulated clusters and we clarify the terminology used.
Out results  are presented in Section~\ref{sec:results}, where we discuss the
relation between the level of HE and the cluster dynamical and thermal
structure, as well as its relation with mass- and temperature-bias.
Finally, we draw our conclusions in Section~\ref{sec:conclusion}.

%

\section{The simulated data-set}\label{sec:sims}
\looseness=-1
\sloppy{The data set used in this work is constituted by a sample of
  29 simulated clusters analyzed at $z=0$. Among them, 24 are massive
  systems with $M_{200}>8\cdot 10^{14}h^{-1}M_\odot$ and 5 are smaller
  objects with $M_{200}$ in the range $1$--$4\cdot
  10^{14}h^{-1}M_\odot$~\cite[][]{planelles2014}.
These clusters have been selected as the most massive haloes residing
at the centre of 29 Lagrangian regions, re-simulated from zoomed
initial conditions~\cite[the same of][]{bonafede2011}, with the
Tree-PM Smoothed-Particle-Hydrodynamics (SPH) code GADGET-3~\cite[][]{springel2005}.
}
The simulations assume a $\Lambda$CDM cosmological model with
$\Omega_{\rm{m}} = 0.24$, $\Omega_{\rm{b}} = 0.04$,
$H_0= 72\,{\rm km\,s}^{-1}\mpc^{-1}$,
$n_{\rm{s}}=0.96$, and $\sigma_8 = 0.8$.
The mass resolution of this set is $m_{\rm{DM}} = 8.47\times10^8 \,
\msun$ for the DM particles, and $m_{\rm{gas}} = 1.53\times10^8\,
\msun$ for the initial gas particle mass.
The Plummer-equivalent softening length for the computation of the
gravitational force is $\epsilon = 3.75\kpch$ for DM and gas
particles, $\epsilon = 2\kpch$ for star and black hole particles at $z=0$.

The version of the code used here includes the improved version of the
hydrodynamical scheme described in~\cite{beck2015}, that largely
improves the SPH capability to follow gas-dynamical instabilities and
mixing processes, and prevents particle clumping. In particular, these
new developments include a higher-order interpolation kernel as well
as time-dependent formulations for artificial viscosity and artificial
thermal diffusion. More details on the hydrodynamical method as well
as a large set of standard tests are presented in~\cite{beck2015}.

\sloppy{The physical processes treated in the simulations comprise
metallicity-dependent radiative cooling, time-dependent UV background,
star formation from a multi-phase inter-stellar
medium~\cite[][]{springel2003}, metal enrichment from supernovae (SN)\,II, SN\,Ia
and asymptotic-giant-branch stars~\cite[][]{tornatore2004,tornatore2007}, SN-driven
kinetic feedback in the form of galactic winds (with $350\,{\rm
  km\,s^{-1}}$ velocity), and the novel model for AGN thermal feedback,
presented in~\cite{steinborn2015}, in which cold and hot gas accretion
onto black holes (BHs) is treated separately. In particular, we consider here only
the cold-phase accretion, assuming $\alpha_{\rm cold}=100$ as boost
factor of the Bondi rate for the Eddington-limited gas accretion onto
the BH~\cite[see also][]{gaspari2015}.}

For this paper, we employ a set of simulations in which all the above
physical processes are included. This allows us to reproduce the ICM
as realistically as possible. This new set of simulations has
been recently presented in~\cite{rasia2015}, where it was shown how
it was possible for the first time to recover the observed coexistence
of cool-core (CC) and
non-cool-core (NCC) clusters~\cite[][]{rasia2015}.
More results on the simulations will also be presented in
forthcoming papers (Murante et al., in prep.; Planelles et al., in
prep.).


\section{Characterizing the deviation from HE}\label{sec:method}
With the use of hydrodynamical simulations it is possible to trace
directly the 3D structure of the gas acceleration field.  In
particular, from the GADGET code we obtain the value of the gas total
acceleration ($d{\bf v}/dt$, Eq.~\eqref{eq:lagr-deriv}) for each gas particle in the
simulation output, explicitly separated in its gravitational and hydrodynamical components.

In order to satisfy hydro-{\it static} equilibrium, the two acceleration components must balance:
\begin{equation}\label{eq:acc}
{\rm HE:}\qquad
0 = \frac{d{\bf v}}{dt}
= {\bf a} = {\bf a}_{g} + {\bf a}_{h}.
\end{equation}

  In general, the equilibrium in Eq.~\eqref{eq:acc} should
  be evaluated separately for each component of the acceleration
  vector.
  However, in case of astronomical objects such as galaxy clusters or stars, the
  condition has to hold radially.
  For this reason, we
  consider only the radial component of the accelerations ${\bf
    a}_{g}$ and ${\bf a}_{h}$, indicated as \gacc\ and \hacc,
  respectively, that we averaged within spherical shells.

%
\subsection{Method}
We investigate deviations from HE by studying the
  deviation from $-1$ of the ratio, $\ratio$, between the radial
  components of gravitational and hydrodynamical accelerations.

To compute radial profiles of the \gacc/\hacc\ term,
two alternative approaches have been be used:
\begin{enumerate}[(i)]
  \item evaluating the \gacc/\hacc\ ratio particle by particle and then averaging over the spherical shell;
  \item building the profiles of the two accelerations
  separately and then computing the ratio of the \gacc\ radial profile to
  the \hacc\ radial profile.
\end{enumerate}
Methods (i) and (ii) are equivalent in the ideal case of a spherical
gas distribution in HE and without in-homogeneities (see
appendix~\ref{app:hyd_sph}).

We note that all the calculations have been done by subtracting a bulk gravitational acceleration,
which in principle can be non negligible.
This is calculated as a mass-weighted mean value within $\rtwo$, considering
all the particle species (i.e. DM, gas and stars).
The mean value of the hydrodynamical
component is not accounted for
because is typically very low.
We verified that for all the 29 main haloes in the sample,
both  acceleration components are indeed very low.

The gas particles used for the calculation
are those in hot phase.
Namely, we remove from the computation the cold gas
($T<3\cdot10^4$\,K), and the multi-phase gas particles which have a
cold mass fraction greater than 10 per cent.

Given the purposes of our investigation, we do not remove any
substructure.


\subsection{Terminology}

We summarize here the meaning of the quantities and assumptions employed.
\begin{enumerate}[(a)]
\item
  {\bf Acceleration term}: this is the total derivative $d{\bf
  v}/dt$ in Eq.~\eqref{eq:euler}, which contains both the pure
  acceleration term $\partial {\bf v}/\partial t$ and the inertia
  term $( {\bf v} \cdot \nabla){\bf v}$; as previously explained we
  will refer to this term in the form of $\ratio$, assuming spherical
  symmetry and considering the radial component of the acceleration
  only.
\item
  {\bf HE}: from Eq.~\eqref{eq:acc}, HE is quantified by $\ratio =
  -1$, with the underlying assumption of spherical symmetry;
\item
  {\bf Deviation from HE}: $$\dHE =\ratio + 1\,;$$
\item
  $\mathbf{\mhe}$: this indicates the hydrostatic mass and implies
  the assumptions of HE, spherical symmetry, {\it and} purely
  thermal nature of the
  pressure.
\item
  {\bf Hydrostatic-mass bias}: $$b_M= (\mhe-\mtrue) / \mtrue \,,$$
  where $\mtrue$ is the total gravitating mass of the system, computed
  summing up all the particle masses (within the considered radius).
\end{enumerate}

\begin{figure*}[tb]
\centering
\includegraphics[width=0.49\textwidth,trim=20 0 0 10,clip]{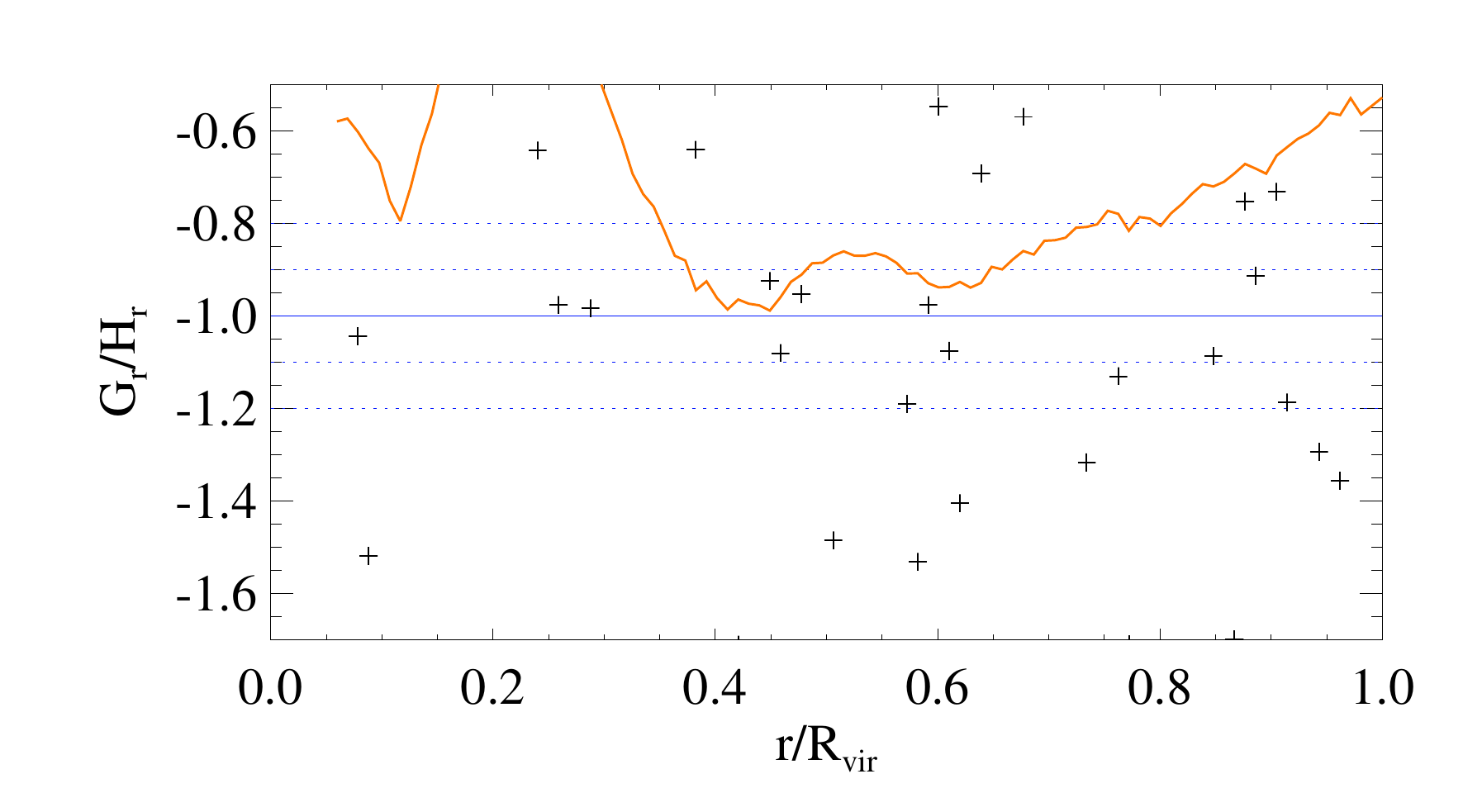}
\includegraphics[width=0.49\textwidth,trim=20 0 0 10,clip]{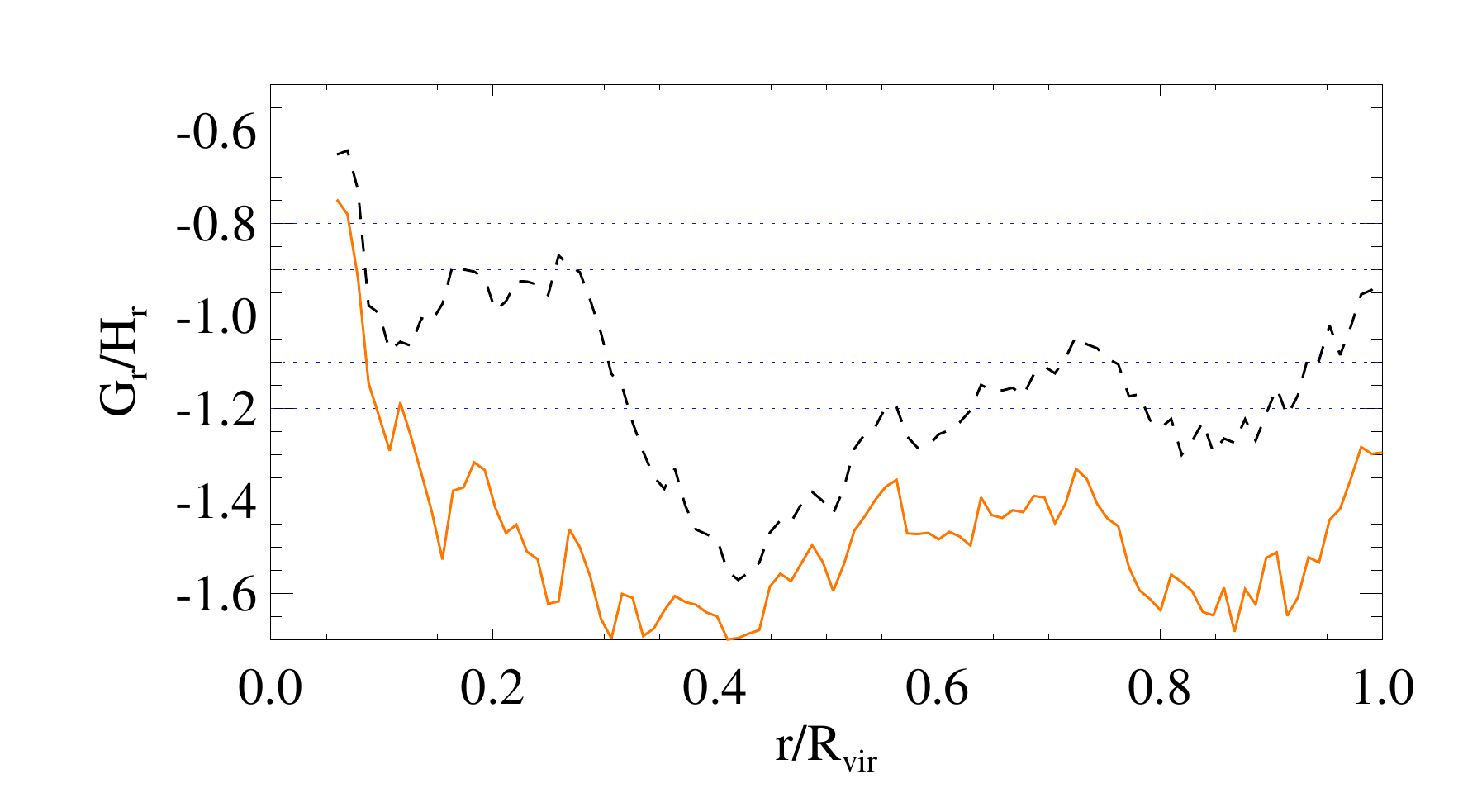}
\caption{Radial profiles of the $\ratio$ ratio for one simulated cluster (D8)
  of the sample. {\it Left:} mean (black crosses) and median (orange,
  solid line) profiles calculated from the particle-by-particle values
  of $\ratio$; {\it right:} mean (black, dashed line) and median
  (orange, solid line) $\ratio$ profiles calculated from the radial
  profiles of the two separate components \gacc\ and \hacc.}
\label{fig:1accprof}
\end{figure*}

It is important to notice that (b) and (d) are derived from the left- and
right-hand-side terms of Eq.~\eqref{eq:euler} {\it but} do not share
the same assumptions, as purely thermal pressure is assumed only when
$\mhe$ is calculated.


\section{Results}\label{sec:results}

\subsection{Application to simulated galaxy clusters}

By applying the method described in the previous section to simulated
galaxy clusters, we can gain a deeper understanding of the intrinsic
state of the ICM, in the presence of several astrophysical phenomena,
such as star formation, feedback processes, and accretion of
substructures.
Moreover, we can explore the 
validity of the HE assumption and its influence on the median
behaviour in different population of clusters.

For the purpose of
showing how complex is the level of HE deviation at different radii,
we start by considering one single object shown in Figure~\ref{fig:1accprof}.

We notice the evident difference
between the
methods $(i)$ and $(ii)$ described in Section~\ref{sec:method}.
In particular,
both the mean (black, dashed curve) and median profiles (orange, solid
curve) are smoother when we measure the ratio of the two profiles
(see right panel in Figure~\ref{fig:1accprof}).

The different picture drawn from the particle-based approach
(left panel in Figure~\ref{fig:1accprof}) can be
ascribed to the inhomogeneous distribution of the gas accelerations
and  to the large spread of the hydrodynamical acceleration
values, which is considerably broader than the gravitational one.
Furthermore, the kernels used to smooth the hydrodynamical and
gravitational forces are different. Therefore, the accelerations are
evaluated at two unequal scales.
Furthermore, numerical terms (e.g.\ artificial viscosity and
diffusion) intervene in the SPH implementation of the Euler
equation, so that Eq.~\eqref{eq:euler} is not satisfied in its
theoretical formulation on a particle base.
In the following we use the second
approach where we compute the two \gacc\ and \hacc\ components separately and
subsequently we calculate the ratio $\ratio$ from their profiles.

\enlargethispage*{\baselineskip}
The cluster shown in Figure~\ref{fig:1accprof} represents a rather extreme case,
with deviations up to $\sim 50$ per cent or more,
already outside the inner region ($>0.1\,\rvir$),
indicating a non-negligible violation of the HE assumption and
possible biases in the HE-derived mass estimate.

In general, among the 29 clusters of the sample, the individual
profiles show significant variations and a close inspection to the
distribution of each cluster substructures, merging and thermal
history would be necessary to understand the detailed features of the
$\ratio$ profiles.


\subsection{On the relation to the hydrostatic mass bias}\label{subsec:sigr}

\begin{figure*}[tb]
\centering
{\bf D5}\\
\includegraphics[width=0.33\textwidth,trim=20 10 20 20,clip]{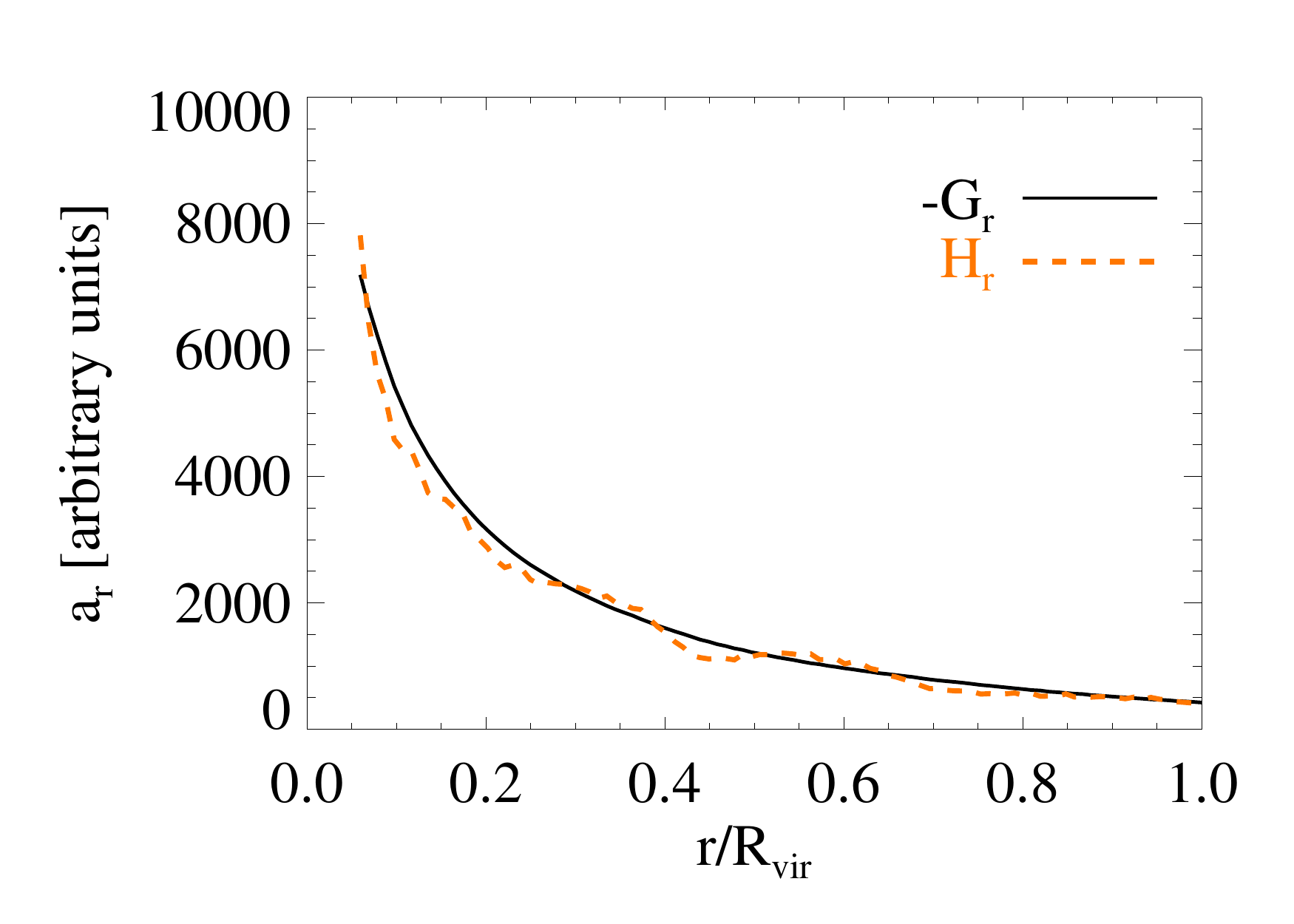}
\includegraphics[width=0.33\textwidth,trim=20 10 20 20,clip]{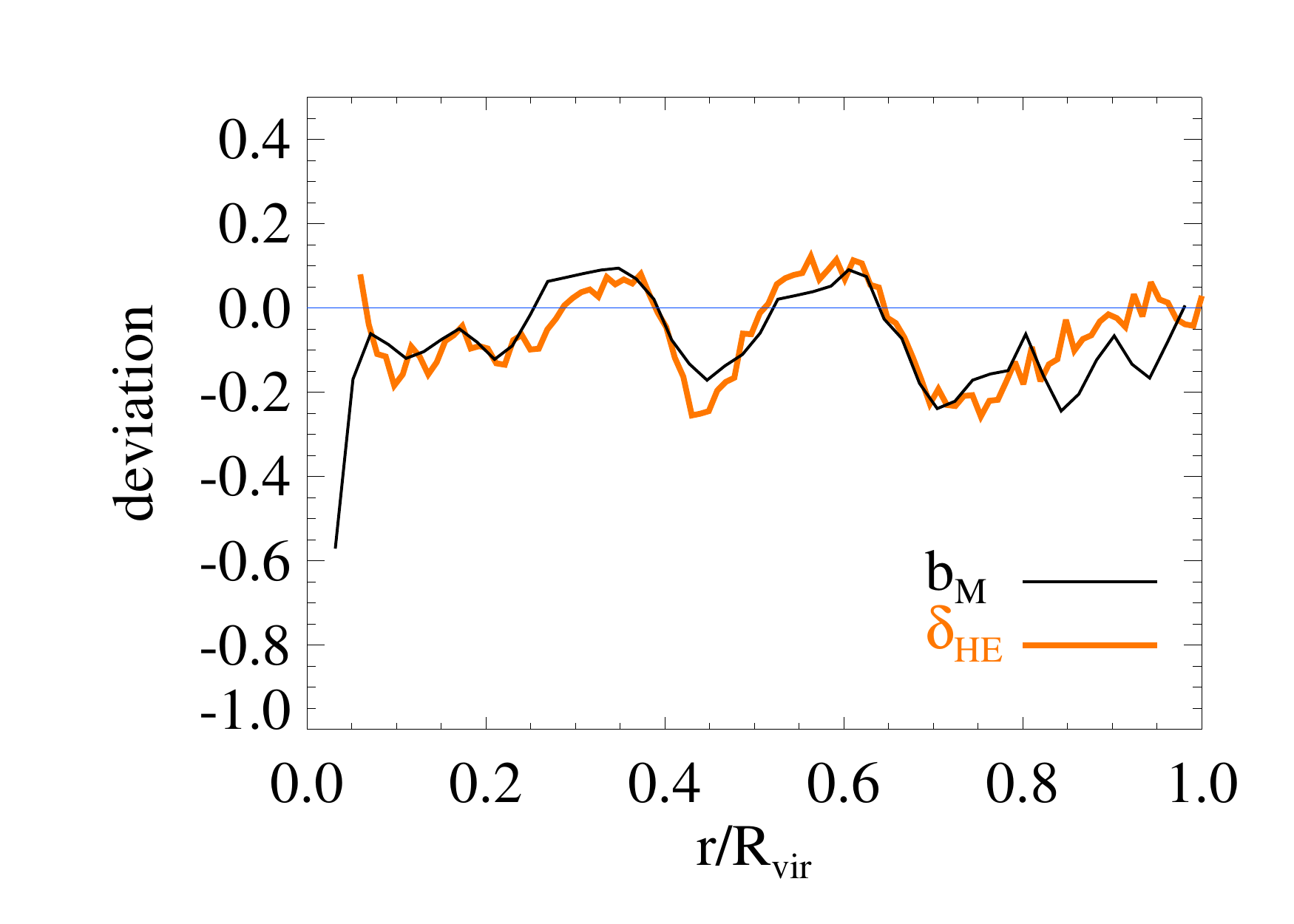}
\includegraphics[width=0.33\textwidth,trim=20 10 20 20,clip]{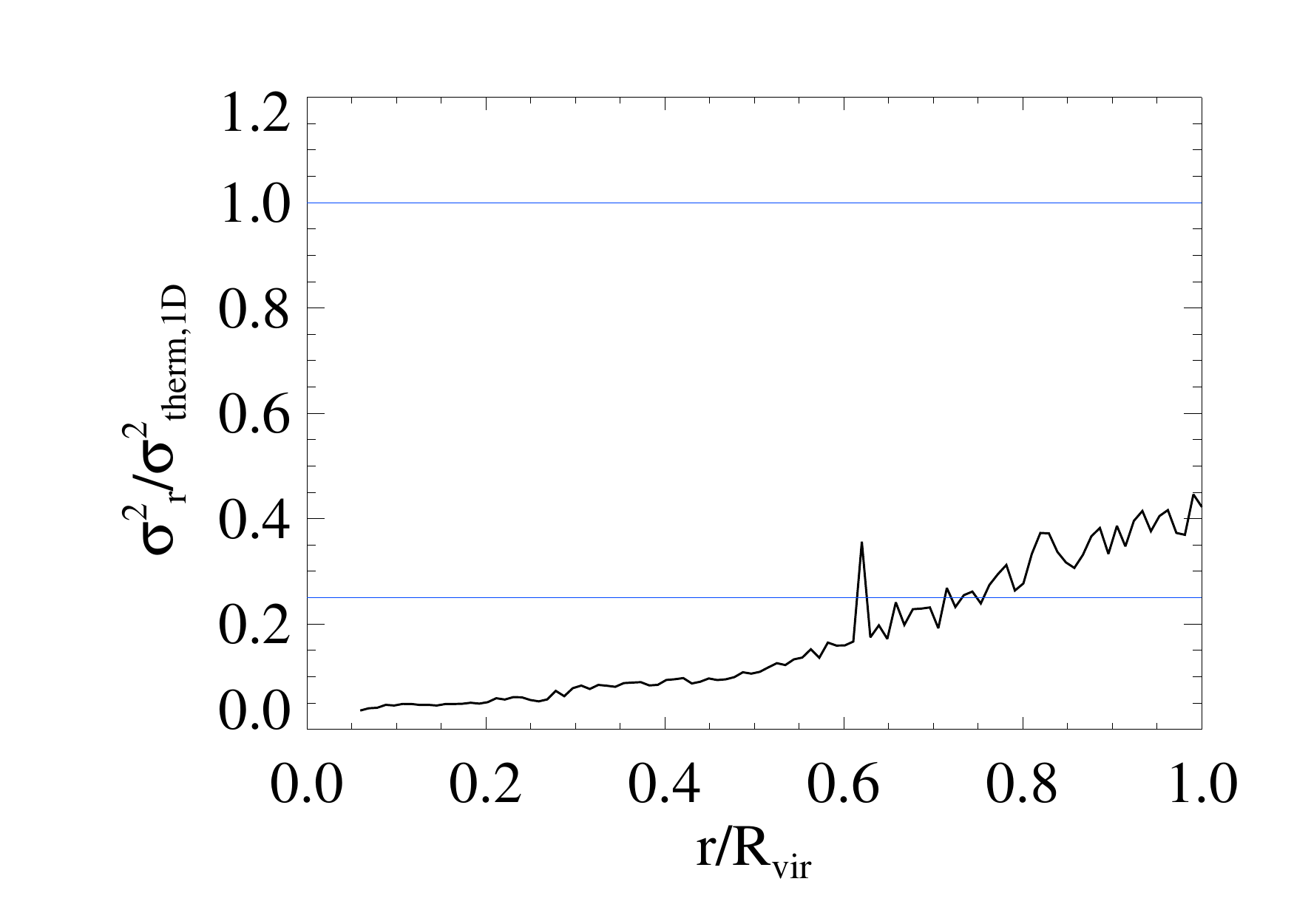}
\\{\bf D24}\\
\includegraphics[width=0.33\textwidth,trim=20 10 20 20,clip]{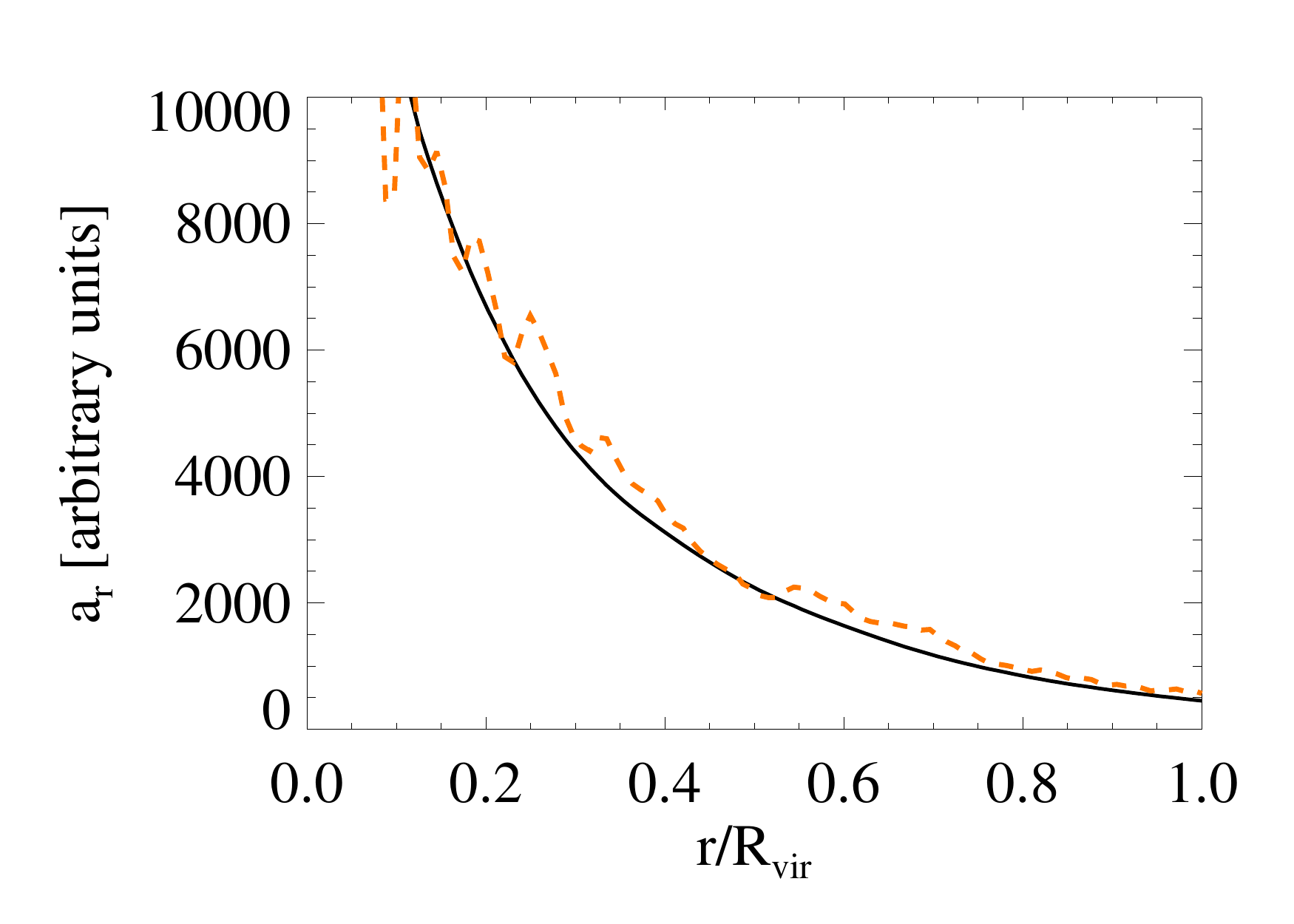}
\includegraphics[width=0.33\textwidth,trim=20 10 20 20,clip]{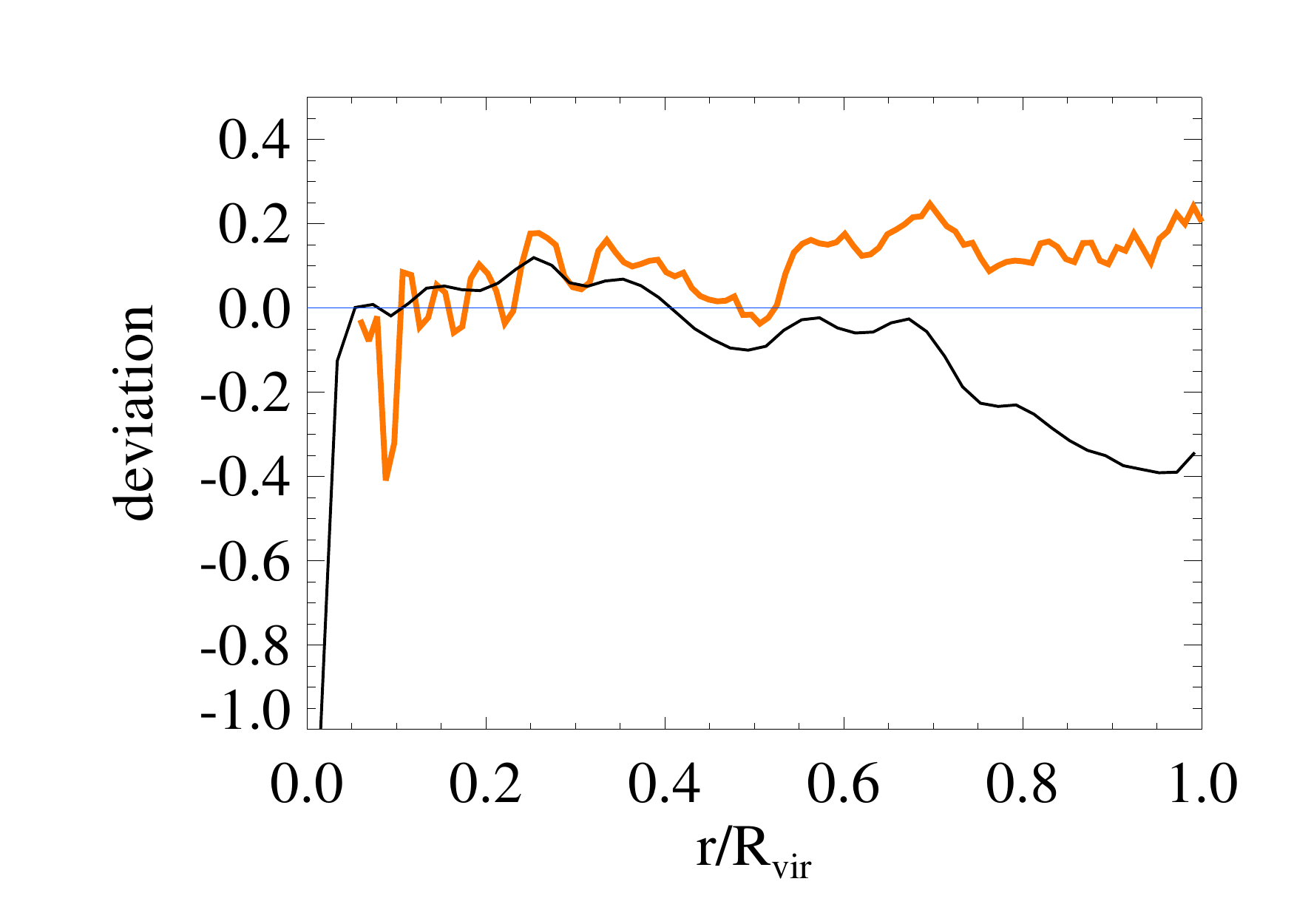}
\includegraphics[width=0.33\textwidth,trim=20 10 20 20,clip]{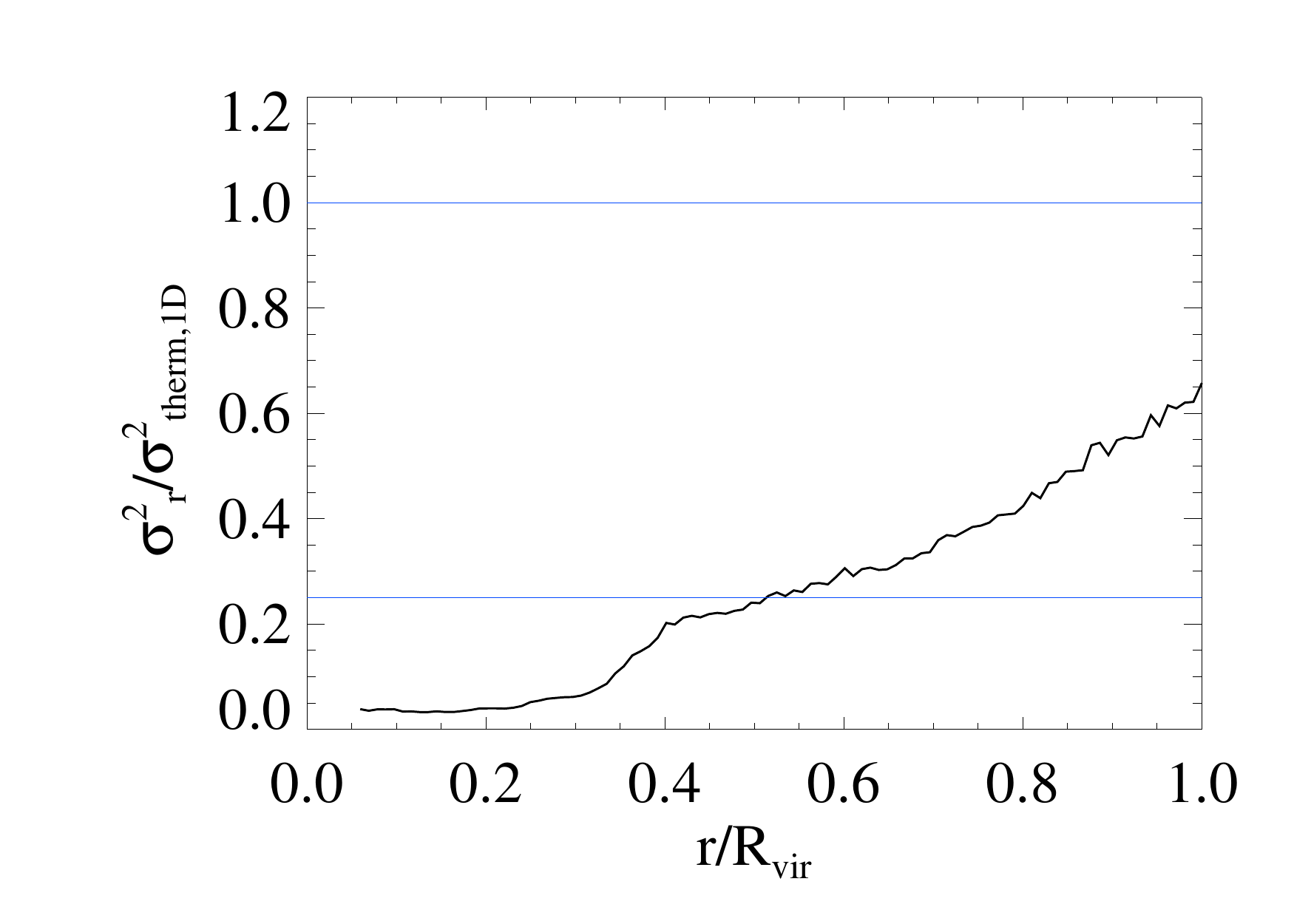}
\\{\bf D8}\\
\includegraphics[width=0.33\textwidth,trim=20 10 20 20,clip]{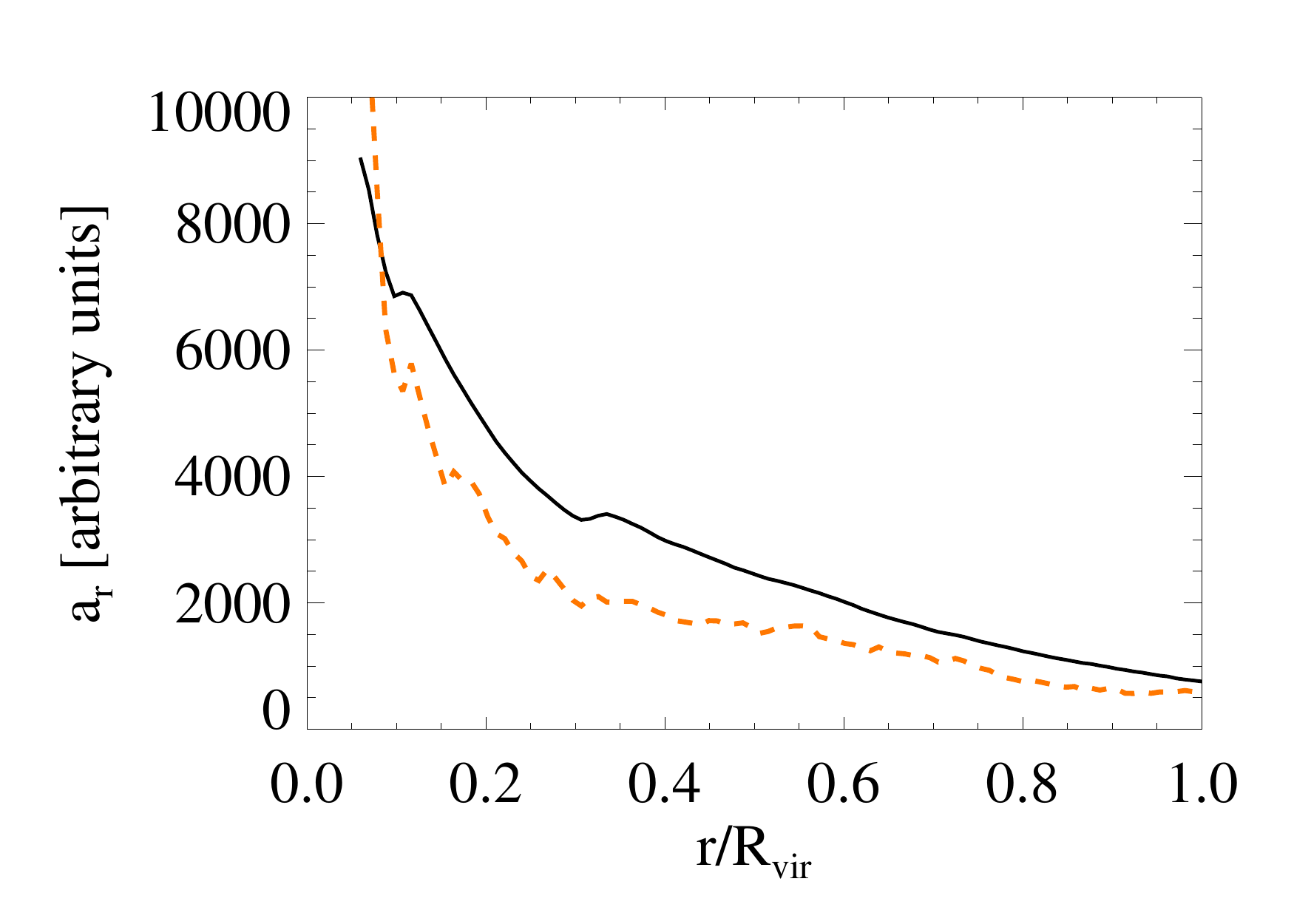}
\includegraphics[width=0.33\textwidth,trim=20 10 20 20,clip]{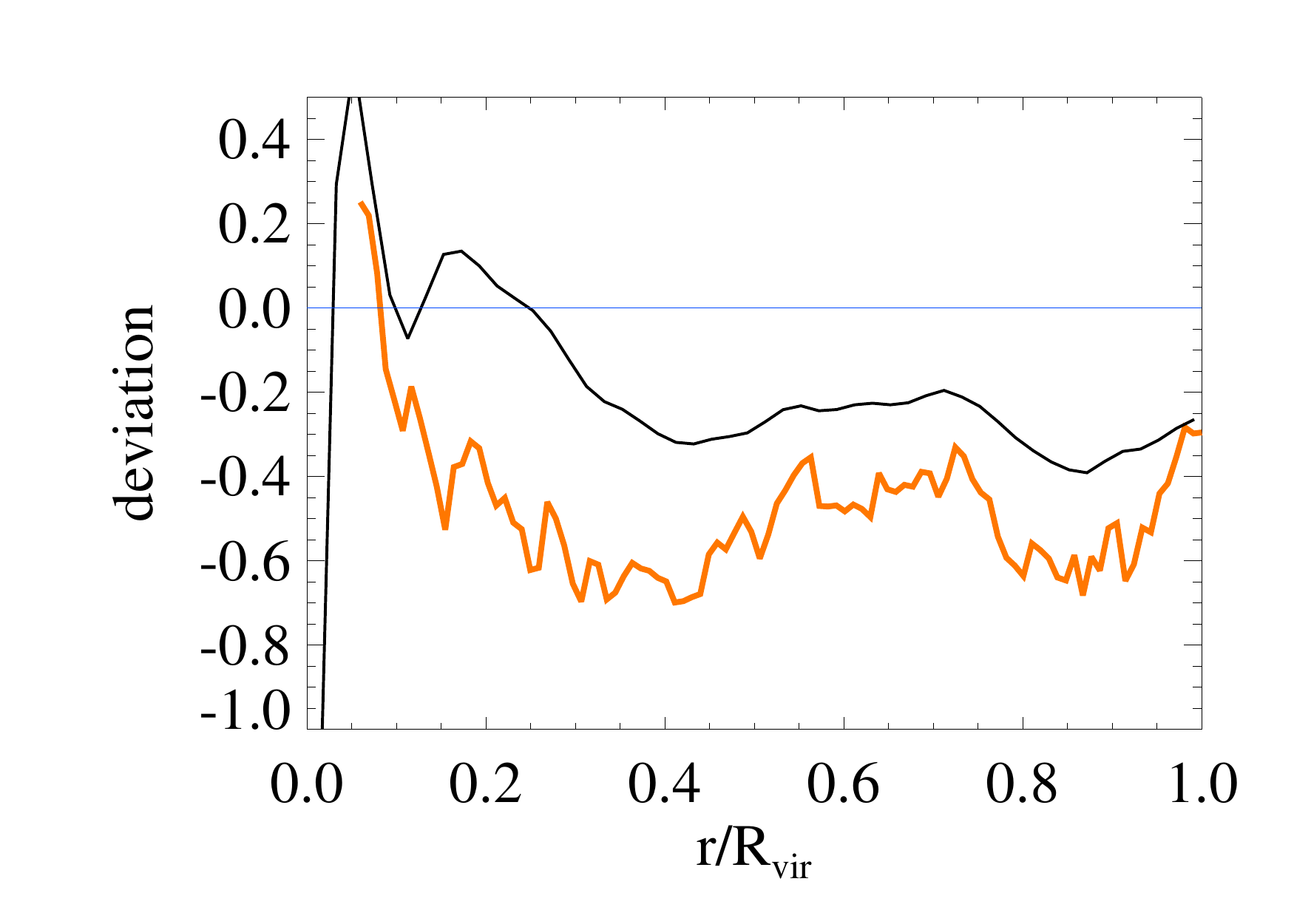}
\includegraphics[width=0.33\textwidth,trim=20 10 20 20,clip]{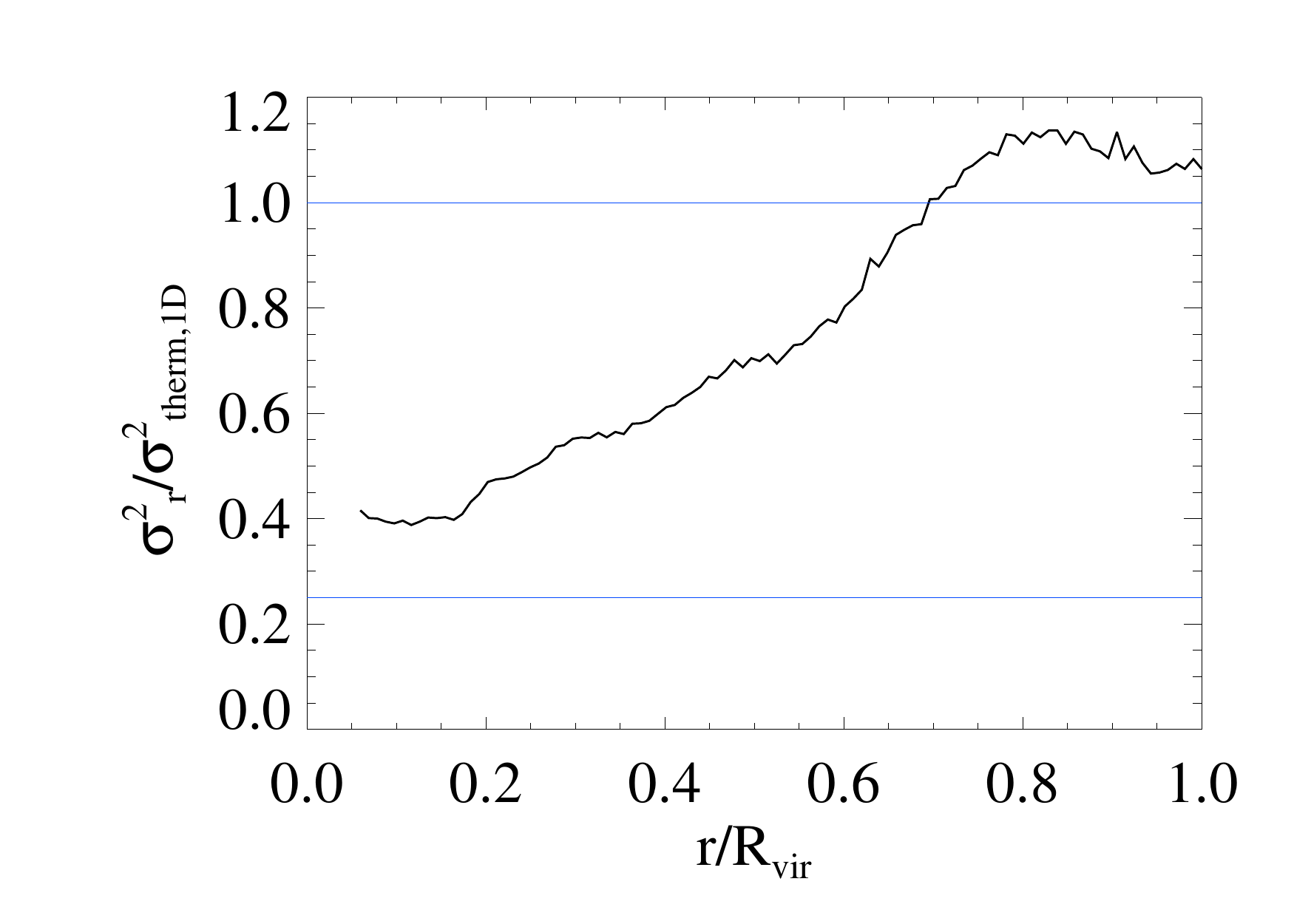}
\caption{Three different clusters (D5, D24 and D8) in the sample,
  chose to represent two extreme cases and an intermediate one
  w.r.t. the comparison between the deviation from HE, traced by $\dHE
  = \ratio + 1$, and mass bias $b_M$.  From left to right we show the
  radial profiles of: the gravitational (\gacc, changed in sign;
  black solid line) and hydrodynamical (\hacc; orange dashed line)
  accelerations (radial components); mass bias ($b_M$; black line) and
  deviation from HE ($\dHE$; orange thick line); amount of non-thermal
  motions along the radial direction, with respect to thermal
  ones.\label{fig:2halo_compar}}
\end{figure*}

For each cluster in the sample we calculate the hydrostatic mass
as in Eq.~\eqref{eq:mhe}. For the purpose of our theoretical
investigation, we do not apply any substructure removal from the ICM.
In principle, the hydrostatic mass bias, $b_M$ (definition (e),
in section~\ref{sec:method}), does quantify the deviation from HE as
long as the additional hypotheses on which the hydrostatic mass
relies are valid.  Therefore, it is interesting to compare the
radial profiles of $\ratio$ and mass bias.

In order to explore the origin of the differences between mass bias
and deviation from HE, it is also useful to investigate possible
connections to the level of non-thermal motions of the gas,
which are typically not accounted for in the usual hydrostatic mass
estimate.

In Figure~\ref{fig:2halo_compar} we present, for three clusters of the
sample, the separate profiles of the two accelerations (left panels),
the direct comparison of mass bias and deviation from HE (central
panels), and the ratio $\sigma^2_r/\sigma^2_{\rm therm,1D}$ (right
panels), where $\sigma^2_r$ is the velocity dispersion of the gas in
the radial direction and $\sigma^2_{\rm therm,1D}$ is the expected
one-dimensional thermal velocity dispersion,\footnote{Here, $\sigma^2_r$ is computed as the dispersion on the radial component of the gas velocities with respect to the mean value, in each radial shell.}
The thermal velocity dispersion, instead,
is calculated~as $$
\sigma^2_{\rm therm} = \frac{3k_B \tmw}{\mu m_p},$$ where $\tmw$ is the
mass weighted temperature in the shell and for a single dimension
$\sigma^2_{\rm therm,1D} = \sigma^2_{\rm therm}/3$.} in the same radial bin.
The last quantity represents the excess of the velocity dispersion
produced by bulk and random motions over that produced by thermal
motions.

\looseness=-1  The top- and bottom-row panels represent the opposite cases where the
  profiles of $b_M$ and $\dHE$ either trace each other in a very
  good way (as for D5) or show a significant off-set (D8).  In the
  first case, the mass bias more directly reflects the level of
  deviation from HE, suggesting that the additional assumption of
  purely thermal pressure support --- included in $\mhe$, but not
  involved in $\dHE$ --- does not play a significant~role. This is in
  fact supported by the low amount of non-thermal motions, with respect
  to thermal ones, shown in the right panel.
  Viceversa for D8, we notice a systematic difference between $b_M$
  and $\dHE$, with the latter much more significant than the mass
  bias, throughout the whole radial range. The origin of the
  significant deviation from HE on the radial direction is due to the
  systematic unbalance between the two forces, as visible from the
  separate \gacc\ and \hacc\ profiles in the left panel.  The mismatch
  between $b_M$ and $\dHE$ is strongly connected with the behavior of
  the $\sigma^2_r/\sigma^2_{\rm therm,1D}$ ratio, which significantly
  increases from 0.4 in the center\footnote{Here and throughout the paper, the cluster center corresponds to the position of the DM particle having the minimum value of the potential.}
 out to more than one towards the
  virial radius, indicating that outside $\sim 0.7 \rvir$ macroscopic
  non-thermal motions actually start to be dominant.

  The halo presented in the middle row of
  Figure~\ref{fig:2halo_compar} (D24) represents instead an
  intermediate case where, despite the low-level deviation from
  $\ratio =-1$ ($\dHE$, mostly around $10$ per cent out to the virial
  radius), the mass bias profile does show a different trend,
  especially towards the outer regions. The modulus of $\ratio$
  decreases with increasing radius, indicating that \hacc\ dominates
  over \gacc,\footnote{As visible from the two acceleration profiles
    in the left panels, the sign of \gacc\ is negative, while the sign
    of \hacc\ is positive.}  whereas the mass bias suggests the
  opposite unbalance between the hydrodynamical and the gravitational
  forces: $b_M<0$, indeed, indicates that the thermal pressure support
  {\it under}-estimates the gravitating mass.

\begin{figure*}[tb]
\centering
\includegraphics[width=0.35\textwidth,trim=20 10 0 10,clip]{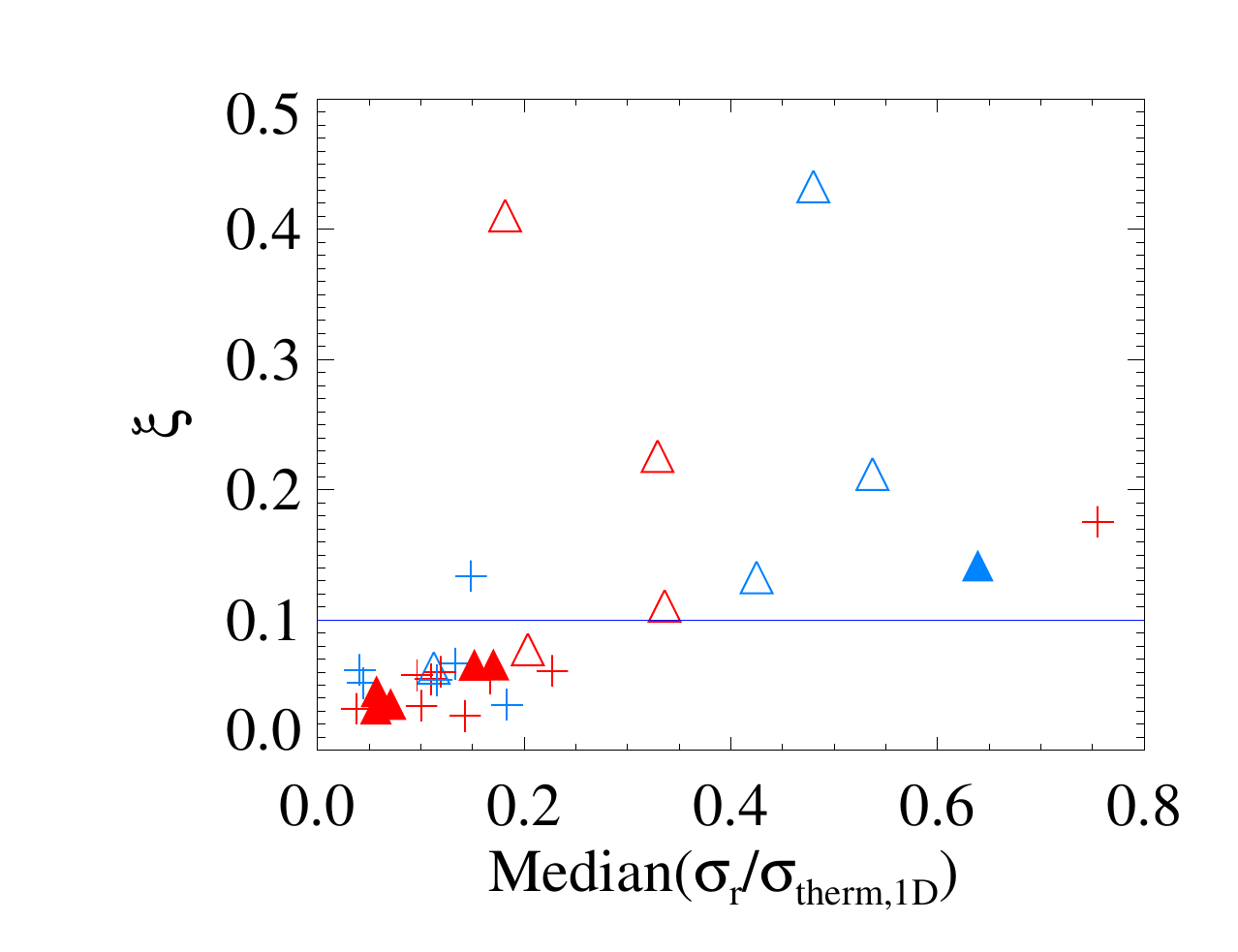}\quad
\includegraphics[width=0.49\textwidth,trim=20 10 0 10,clip]{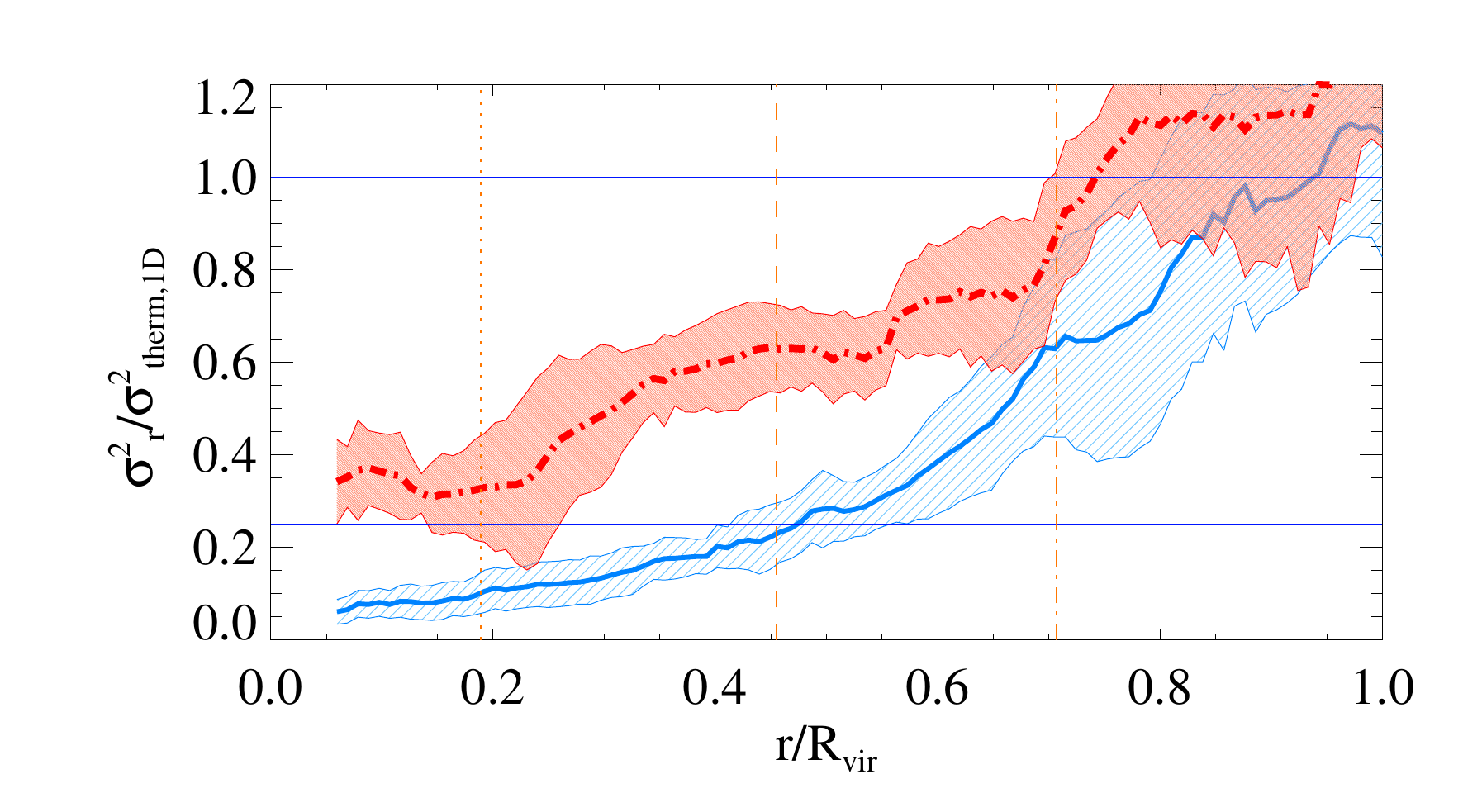}
\caption{{\it Left:} relation between the median difference
  between the profiles of mass bias and deviation from HE (quantified
  by $\xi$; see Eq.~\eqref{eq:mu}), and the median value of the
  $\sigma^2_r/\sigma^2_{\rm therm,1D}$ ratio, within $\rfive$, for the
  29 clusters in the sample. The horizontal line marks the threshold
  used to identify the subsample of clusters with large(small)
  differences between the $b_M$ and $\dHE$ profiles, i.e. with
  $\xi>0.1$($<0.1$).
The symobls and colors refer to the two classifications used to subdivide the sample, as specified in the following Section~\ref{sec:class}.
  {\it Right:} median radial profile of the
  $\sigma^2_r/\sigma^2_{\rm therm,1D}$ ratio, distinguishing between
  two classes of clusters, i.e.\ those with $\xi<0.1$ (blue solid
  line) and $\xi>0.1$ (red dot-dashed line), within $\rfive$. Shaded
  areas mark the median absolute deviation in each radial bin. From
  left to right, vertical lines mark median values of $\rtwofive$,
  $\rfive$ and $\rtwo$, respectively.\label{fig:sigr-med}}
\end{figure*}

  The trend shown by the three examples in
  Figure~\ref{fig:2halo_compar} is in fact present in the entire
  sample, with large deviations from HE generally associated to
  substantial non-thermal motions of the gas.
  Furthermore, we notice that the
  profiles of \gacc\ and \hacc\ (left panels) indicate that the
  gravitational component is typically smoother than the
  hydrodynamical one and that haloes with large deviations from HE
  also show a large offset of \hacc\ with respect to \gacc, with the
  latter generally dominating in modulus.
  In general, the origin of deviations from HE and its relation to the
  mass bias can significantly vary from cluster to cluster, requiring
  a dedicated investigation of the particular cluster properties.
  Nonetheless,
  from the analysis of all the 29 clusters, we can conclude
  that large differences between $b_M$ and $\dHE$ typically correspond
  to $\sigma^2_r \gtrsim 0.3\,\sigma^2_{\rm therm,1D}$, as observed for
  example in the extreme case of cluster D8, where
  $\sigma^2_r/\sigma^2_{\rm therm,1D}$ is larger than 40 per cent from the
  center out to the virial radius.

From a more quantitative perspective, we provide a measure of the
typical difference between the two radial profiles of $b_M$ and $\dHE$
as the median value of the absolute difference among the two, i.e.
\begin{equation}\label{eq:mu}
\xi = {\rm Median}(| b_M-\dHE |),
\end{equation}
considering the radial range up to $\rfive$.
The motivation to consider the region enclosed by $\rfive$ is that
this is an optimal region targeted also by observational analyses,
while outer portions of the cluster regions are generally more
difficult to characterize. Furthermore, the amount of non-thermal
motions in the radial direction generally increases towards the
outskirts, where merging and accretion processes
play a more significant role, for all the clusters.
Also, the use of the median deviation
allows us to estimate the typical difference between the two profiles,
without being biased by large differences restricted to few radial
bins.  As visible from the left panel in Figure~\ref{fig:sigr-med}, it
is possible to identify a subsample of clusters for which the median
difference between the $b_M$ and $\dHE$ radial profiles, $\xi$, and
the median value of $\sigma^2_r/\sigma^2_{\rm therm,1D}$, are both
very low within $\rfive$. From this figure, we use the threshold
$\xi=0.1$ to separate these systems from those with larger values of
both the indicators.  The median radial profile of
$\sigma^2_r/\sigma^2_{\rm therm,1D}$ for these two subsamples of
clusters is shown in the r.h.s.\ panel of Figure~\ref{fig:sigr-med},
where the red, dot-dashed curve represents the subsample with
typically different $b_M$ and $\dHE$ profiles, while the blue, solid
line indicates those with more similar ones ($\xi<0.1$).  The
comparison of the two profiles confirms that larger differences
between the $b_M$ and $\dHE$ profiles (red curve) correspond to larger
($>30\%$) amount of non-thermal over thermal motions (in the radial
direction), despite the larger dispersion\footnote{Here and throughout
  the paper, we quantify the scatter around the median profile through
  the median absolute deviation, defined as $${\rm M.A.D. = Median(
    x_i - Median(x_i))},$$ with $x_i$ representing the values of the
  individual profiles in every radial bin.} in the subsample that only
comprises 9 systems out of 29. We additionally note that the
systematic offset visible out to $\rfive$, that is in the region where
the classification criteria are defined (Figure~\ref{fig:sigr-med},
left panel), is still present almost out to $\rtwo$ ($\sim
0.7\rvir$). Both subsamples, instead, behave more similarly in the
outermost region, where in fact the $\sigma^2_r/\sigma^2_{\rm
  therm,1D}$ ratio increases for both classes.

  As displayed by Figure~\ref{fig:sigr-med-class} below, a similar behaviour
  is recognised when the subsample is instead divided in regular and
  disturbed systems, with the second class typically showing a higher
  profile of the non-thermal to thermal motion ratio.



\subsection{Average deviation from HE}
\looseness=-1
Overall, our results confirm that
not only the hydrostatic mass bias, but
also the level of deviation from the static assumption, $\dHE$, is
in fact very different from cluster to cluster and at different radii
within the single object.  This is consistent with the
findings obtained in similar works based on AMR simulations
\cite[e.g.][]{shi2015,nelson2014a,lau2013}.

A possible way to investigate the problem is to consider samples of
clusters and stack the individual profiles to infer an average
behaviour.  In this way,
the effects due to asphericity of the individual clusters are
alleviated and
the importance of the assumption of spherical symmetry is less
significant.

\begin{figure}[tb]
\centering
\includegraphics[width=0.47\textwidth,trim=20 7 0 20,clip]{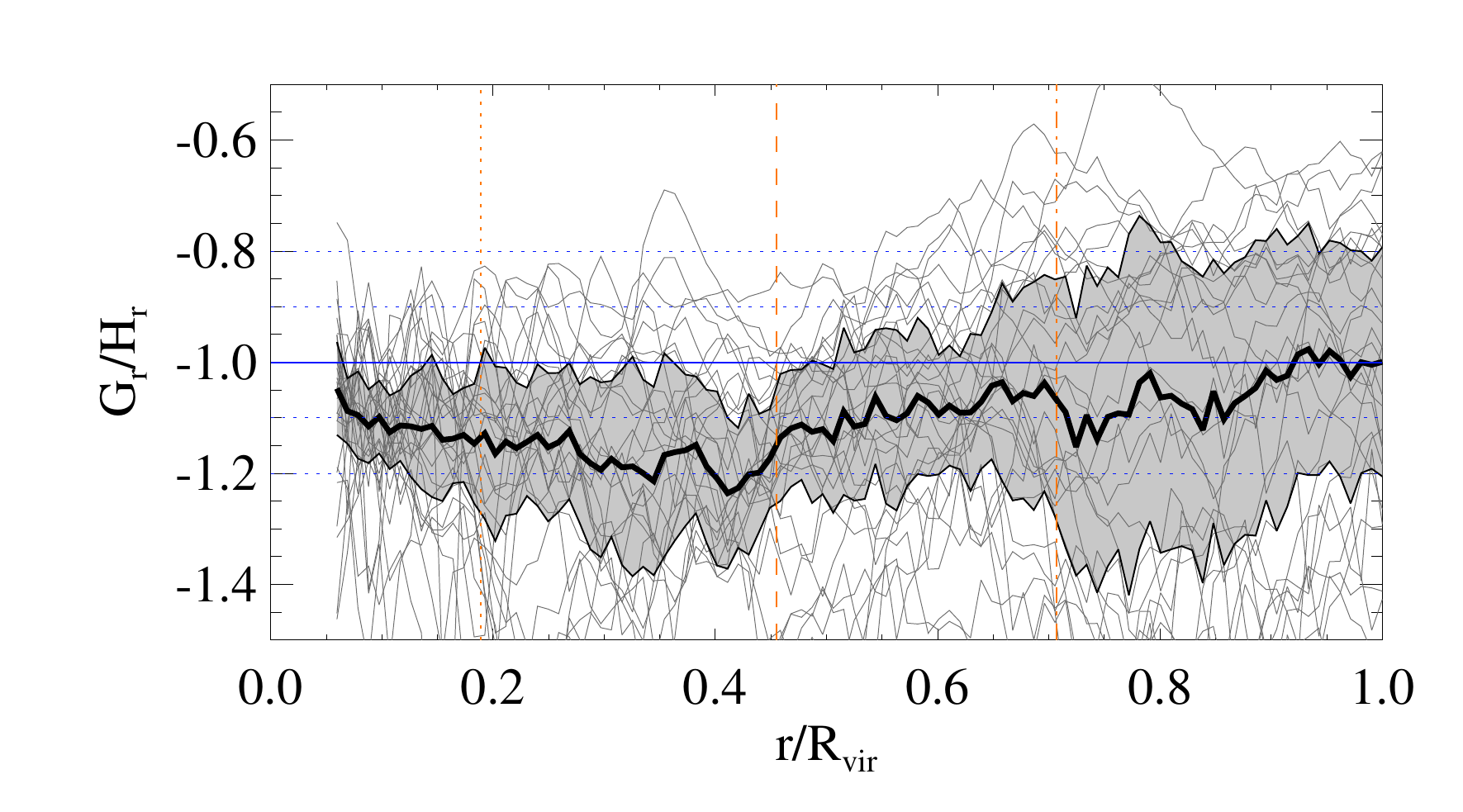}
\caption{Median radial profile (black solid curve) of the ratio
  between the radial components of gravitational and hydrodynamical
  acceleration, for the main haloes in the 29 re-simulated regions.
  The light-grey lines in the background represent the individual
  $\ratio$ profiles of all the 29 haloes.  From left to right,
  vertical lines mark median values of $\rtwofive$, $\rfive$ and
  $\rtwo$, respectively. We indicate with the shaded area
  the median absolute deviation from the
  the median profile.\label{fig:acc-med-all}}
\end{figure}
\looseness=-1 The stack analysis is shown in
Figure~\ref{fig:acc-med-all}, where we display the individual profiles
(grey curves) and the median one (black). The shaded area indicates
the dispersion of the distribution in each radial bin around the
median value, and is computed as the median absolute deviation.
As pointed out, the background individual profiles show very different
features and the overall dispersion increases with the radius.
In the outskirts, where the gas acceleration field is more sensitive to
substructure infalling onto the main~halo, the spread is larger.
Nevertheless, the median profile indicates that the typical deviation
of $\ratio$ from $-1$ is $\sim 10$~per cent, reaching $\sim
20$~per cent at most.

\begin{figure}[tb]
\centering
\includegraphics[width=0.49\textwidth,trim=20 10 0 10,clip]{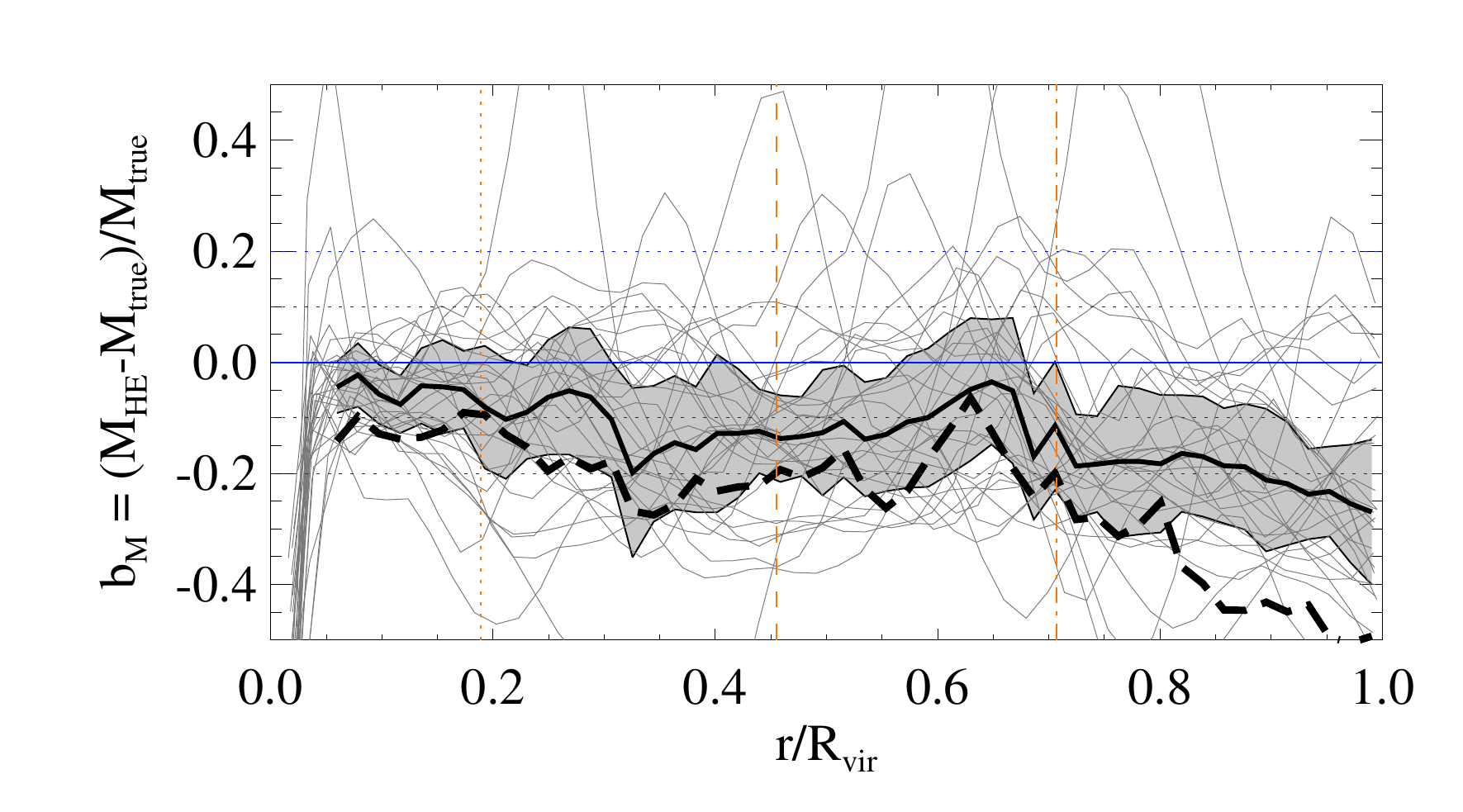}
\caption{Median radial profile of the mass bias $b_M$, with $\mhe$
  calculated for both $T=\tmw$ (black solid line) and $T=\tsl$ (black
  dashed line). Only for the first case ($T=\tmw$), we also report the
  individual mass-bias profiles of the 29 clusters (light-grey curves)
  and the median absolute deviation from the
  median profile (shaded area).  From
  left to right, vertical lines mark median values of $\rtwofive$,
  $\rfive$ and $\rtwo$, respectively.\label{fig:mbias-med}}
\end{figure}

Similarly, we can construct the median profile of the mass bias for
the 29 haloes in the sample, as shown in Figure~\ref{fig:mbias-med}.%
Without distinguishing the dynamical state of clusters, from the
comparison between Figures~\ref{fig:acc-med-all}
and~\ref{fig:mbias-med} we see that the two median profiles of
acceleration term and hydrostatic mass bias are quite similar out to
$\sim 0.7\,\rvir$ (roughly $\sim\rtwo$).  This suggests that the
negative mass bias does trace --- on average --- the violation of HE,
i.e. the equilibrium is not static and the (total) acceleration term
is non~zero.

The computation of the hydrostatic mass as in Eq.~\eqref{eq:mhe} from
X-ray data can include an additional bias due to the under-estimate of
the X-ray temperature with respect to the dynamical temperature.
 A good approximation for the temperature measured by Chandra and
 XMM-Newton telescopes is provided by the so-called spectroscopic-like
 temperature $\tsl$~\cite[][]{mazzotta2004}, which is commonly used in
 numerical simulations and it is defined as
\be\label{eq:tsl}
\tsl = \frac{\Sigma_i w_i T_i}{\Sigma_i w_i}\qquad \mbox{with}\quad
w_i=m_i\rho_i T_i^{-3/4}\, .
\ee
In Eq.~\eqref{eq:tsl} $m_i,\rho_i,T_i$ indicate the mass, density and
temperature of the single gas element in the simulation.\footnote{To
  calculate the spectroscopic-like temperature all particles with
  temperature below $0.3\,\kev$ have been discarded.}  It has been
shown in previous works~\cite[e.g.][]{biffi2014} that due to the
presence of inhomogeneities in the X-ray emitting gas there is a
systematic difference between the spectroscopic-like estimate and the
mass-weighted temperature ($\tmw$) that would introduce an additional
bias in $\mhe$~\cite[e.g.][]{rasia2014}.
The origin of such bias is however independent of the assumptions of
HE or purely thermal pressure support, and only depends of the degree
of thermal complexity of the~ICM.

When the spectroscopic-like temperature $\tsl$ is adopted, instead of
$\tmw$, we observe the additional bias (black dashed curve in
Figure~\ref{fig:mbias-med}) due to the different temperature
estimation.
In this case the average bias ranges from $\sim 15$ per
cent to $\sim 25$ per cent within $\rtwo$, while it is more
significant in the cluster outer regions, increasing up to 50 per cent.

Focusing on the outskirts, the trend of the average mass bias profile
is different from the $\ratio$ one. Given the likely presence
of infalling substructures, the hydrostatic estimate is on average
significantly lower than the true mass, although the deviation from
HE (quantified via $\ratio$) is typically closer to zero.
This suggests that at such large radii the origin of
the mass bias is mostly related to the additional assumption
of the purely thermal nature of the
ICM pressure, involved in the computation of $M_{HE}$,
rather than to a pure violation of the static state $d{\bf v}/dt=0$.
Nevertheless, we remind that $\ratio \sim -1$ only asserts the
equilibrium in the radial direction, and differences in the anisotropy
of the two acceleration components can also play a role, especially
in the outskirts.


\subsection{Distinguishing among cluster populations}
\label{sec:class}
\looseness=-1 We proceed to evaluate the median mass bias and the
$\dHE$ profiles for subsamples defined on the basis of either their
thermal or dynamical processes.
In particular, we consider here two classifications:
(a) one linked to the cool-coreness of the object and (b) the other to its
global dynamical state.  These two classifications are typical
ways of distinguishing regular/disturbed clusters in observational (method a)
and numerical (method b) studies.

\paragraph{Cool-coreness.}
The classification (a) is based on the core thermal properties of the
clusters, and specifically on the central entropy value.
In more detail, we define the cluster as cool core (CC) if
\begin{equation}
{\rm CC}:\left\{
\begin{aligned}
K_0 & < 60 \kev/{\rm cm}^2\\
\sigma & < 0.55
\end{aligned}
\right.\label{eq:CC-Ncc}
\end{equation}
and non cool core (NCC) otherwise \cite[see][for more details on this classification]{rasia2015}.
In Eq.~\eqref{eq:CC-Ncc} $K_0$ is the central entropy derived from the
fit of the cluster entropy profile, and $\sigma$ is the
pseudo-entropy, defined as $\sigma = (T_{\rm IN}/T_{\rm OUT}) * ({\rm
  EM}_{\rm IN}/{\rm EM}_{\rm OUT})^{-1/3}$, with the temperature (T)
and Emission Measure (EM) computed within the ``IN'' and ``OUT''
regions, corresponding to $R<0.05\,R_{\rm 180}$ and $0.05\,R_{\rm 180}
< R < 0.2\,R_{\rm 180}$, respectively \cite[see,
  e.g.,][]{rossetti2011}.

With this method we classify 11 clusters out of 29 as CC, and the remaining 18 haloes as NCC.

\paragraph{Dynamical state.}
The method (b), instead, combines two
criteria commonly used in numerical simulations to classify a cluster
as dynamically regular or disturbed: the center shift ($\delta r$), defined as the
spatial separation between the position of the minimum of the
potential and the center of mass, and the fraction of mass associated
to substructures ($f_{\rm sub}$). In our work, we define regular
clusters those for which
\begin{equation}
\left\{
\begin{aligned}
\delta r =~ &|| {\bf x}_{\rm min} - {\bf x}_{\rm cm} || / \rtwo &<& ~0.07& \\
f_{\rm sub} =~& \frac{M_{\rm tot, sub}}{M_{\rm tot}} &<& ~0.1&
\end{aligned}
\right. \, ,\label{eq:reg-dist}
\end{equation}
where ${\bf x}_{\rm min}$ and ${\bf x}_{\rm cm}$ are, respectively, the
position of the minimum of the potential and the center of mass,
$M_{\rm tot}$ is the total mass and $M_{\rm tot, sub}$ is the total
mass in substructures.  For values of $\delta r$ and $f_{\rm sub}$
above those thresholds the clusters are classified as
disturbed. Similar conditions are adopted in \cite{neto2007} and
\cite{meneghetti2014}.  Those systems for which the two criteria
in~\eqref{eq:reg-dist} are not simultaneously satisfied are classified
as intermediate systems.  This second classification defines the state
of the cluster on more global scales, with the quantities above
calculated for each cluster within $\rtwo$.

With this method we split the sample of
  29 clusters into 6 regular and 8 disturbed systems, and 15
  intermediate cases.


\subsubsection{The ratio $\ratio$}

\begin{figure}[tb]
\centering
\includegraphics[width=0.47\textwidth,trim=25 0 10 0,clip]{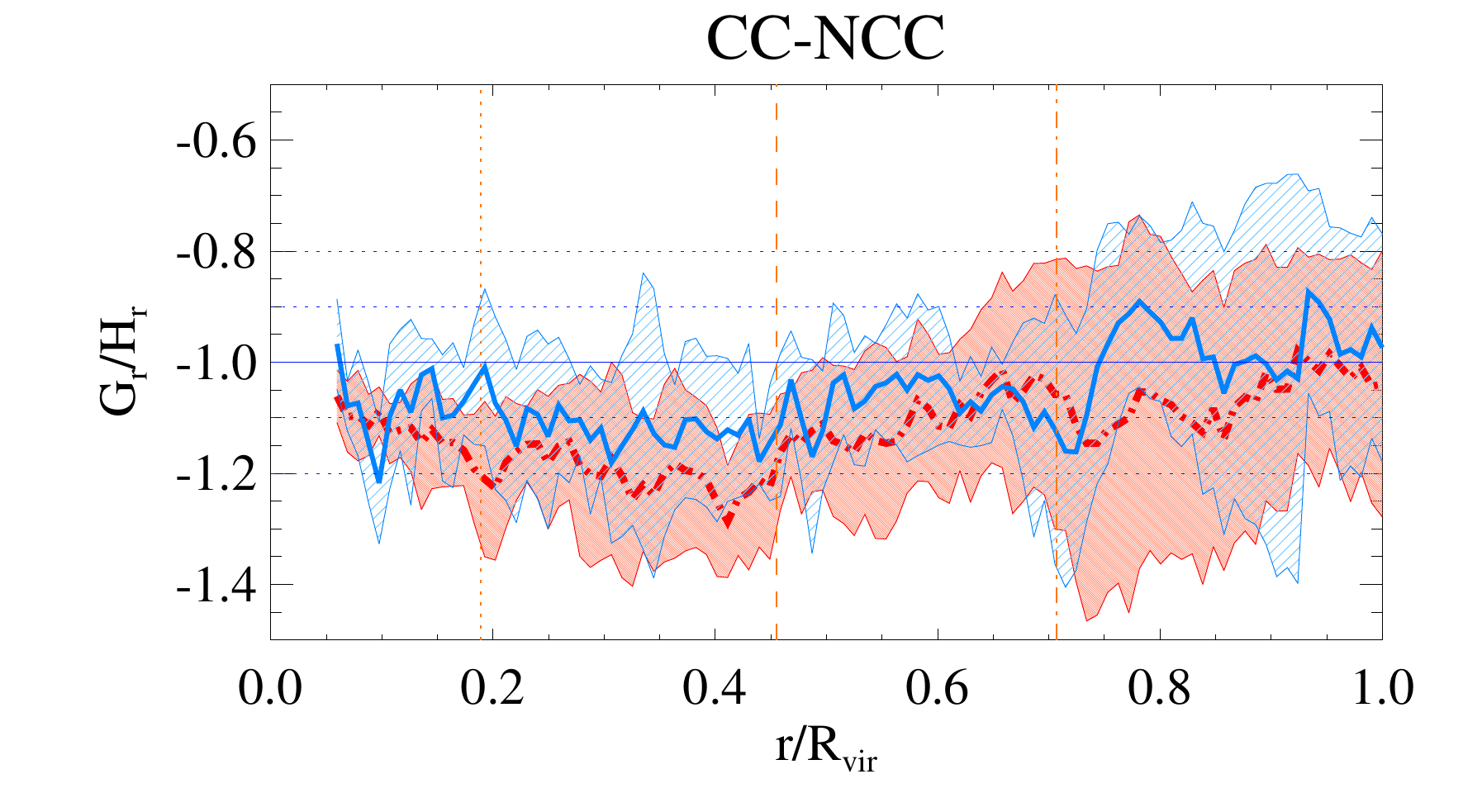}
\includegraphics[width=0.47\textwidth,trim=25 0 10 0,clip]{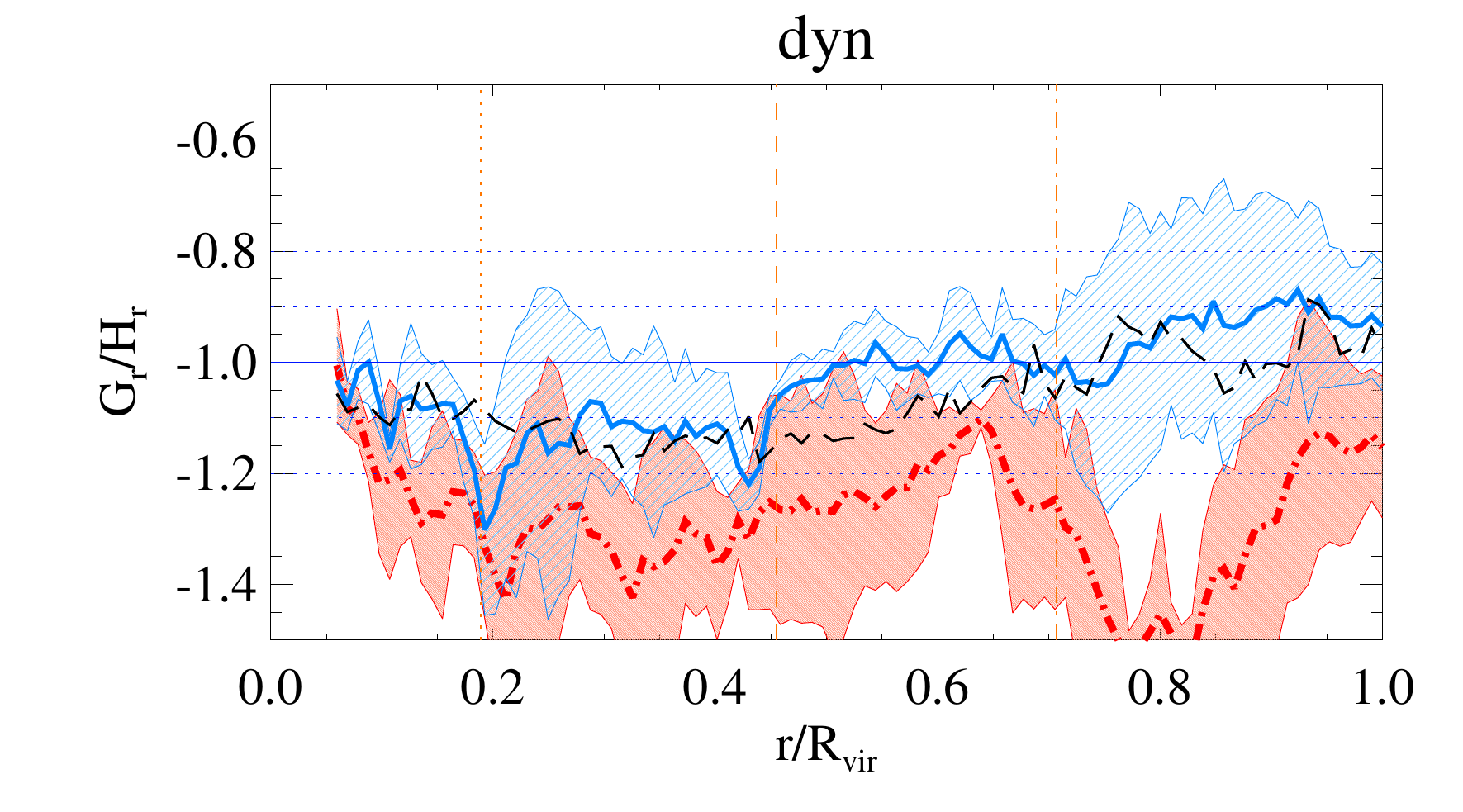}
\caption{Median radial profile of $\ratio$, as in
  Figure~\ref{fig:acc-med-all}, distinguishing among different cluster
  populations.  Shaded areas represent the median absolute
  deviation in each radial bin (w.r.t. the median value). {\it Upper panel:}
  CC/NCC (blue solid/red dot-dashed line); {\it lower panel:}
  regular/disturbed (blue solid/red dot-dashed line) clusters;
  intermediate systems are marked by the thin black line and, for
  simplicity, no dispersion is marked.  From left to right, vertical
  lines mark median values of $\rtwofive$, $\rfive$ and $\rtwo$,
  respectively.\label{fig:acc-med-class}}
\end{figure}

In the two panels of Figure~\ref{fig:acc-med-class} we show that the
median profile of $\ratio$ depends on the classification assumed.

When the sample is divided into CC and NCC clusters, as in the
upper panel of Figure~\ref{fig:acc-med-class}, there is no significant
difference in the $\ratio$ profile of the two populations, especially
considering the dispersion around the median values.  Overall, both
behaviours are very similar to the median profile constructed from the
whole~sample (Figure~\ref{fig:acc-med-all}).

A different picture emerges when the selection is made on the global
dynamical properties. In this case (lower panel of
Figure~\ref{fig:acc-med-class}) the two populations show a clearer
sistematic offset,
especially outside $\rtwofive$,
with the largest departure in
the region outside $\rtwo$ ($\sim 0.7$--$0.9\,\rvir$).
We find that the median profile of regular clusters is
systematically higher and closer to the HE value of
$-1$. On the contrary disturbed clusters show a
larger ($>20$~per cent) deviation from~HE throughout the radial range.

We notice that this difference is not due to the presence of an
intemediate class in the dynamical classification, which has no
analogue in the CC-NCC one. In fact, by restricting the NCC subsample
to the most extreme cases and thus introducing an intermediate class
of objects, a similar separation to that observed between regular and
disturbed clusters is still not found.

It is important to note that our sample of disturbed clusters are
likely to be more strongly affected by merging events and accretion of
infalling substructures, as confirmed also by their clumpiness
profiles, presented by Planelles et al.\ (in prep.).
This should also reflect into a more significant difference in the
(an)isotropy of the gravitational and hydrodynamical acceleration
fields, likely to be enhanced at larger distances from the cluster
center where the mass assembly is still ongoing.

\begin{figure}[tb]
\centering
\includegraphics[width=0.47\textwidth,trim=30 0 10 0,clip]{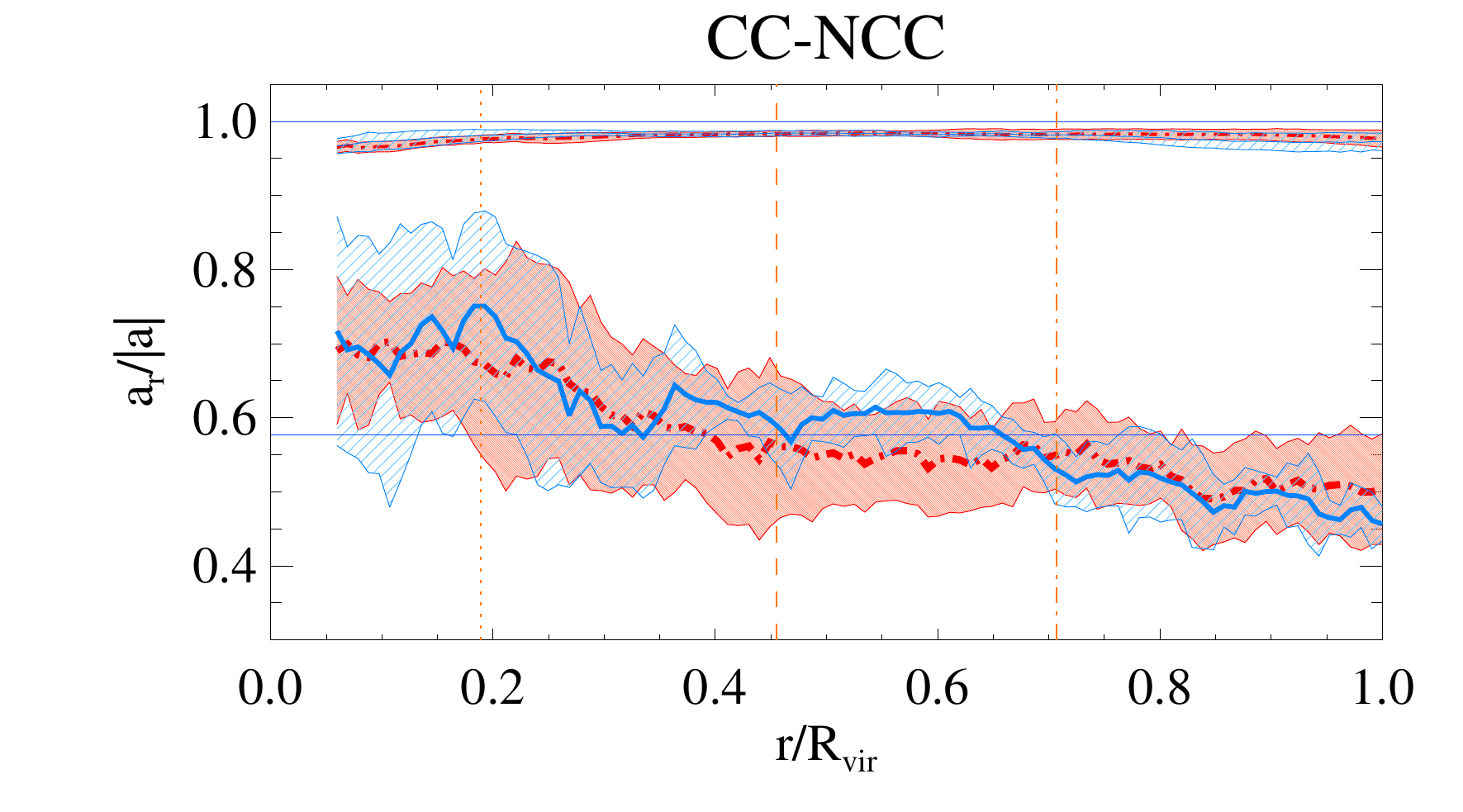}
\includegraphics[width=0.47\textwidth,trim=30 0 10 0,clip]{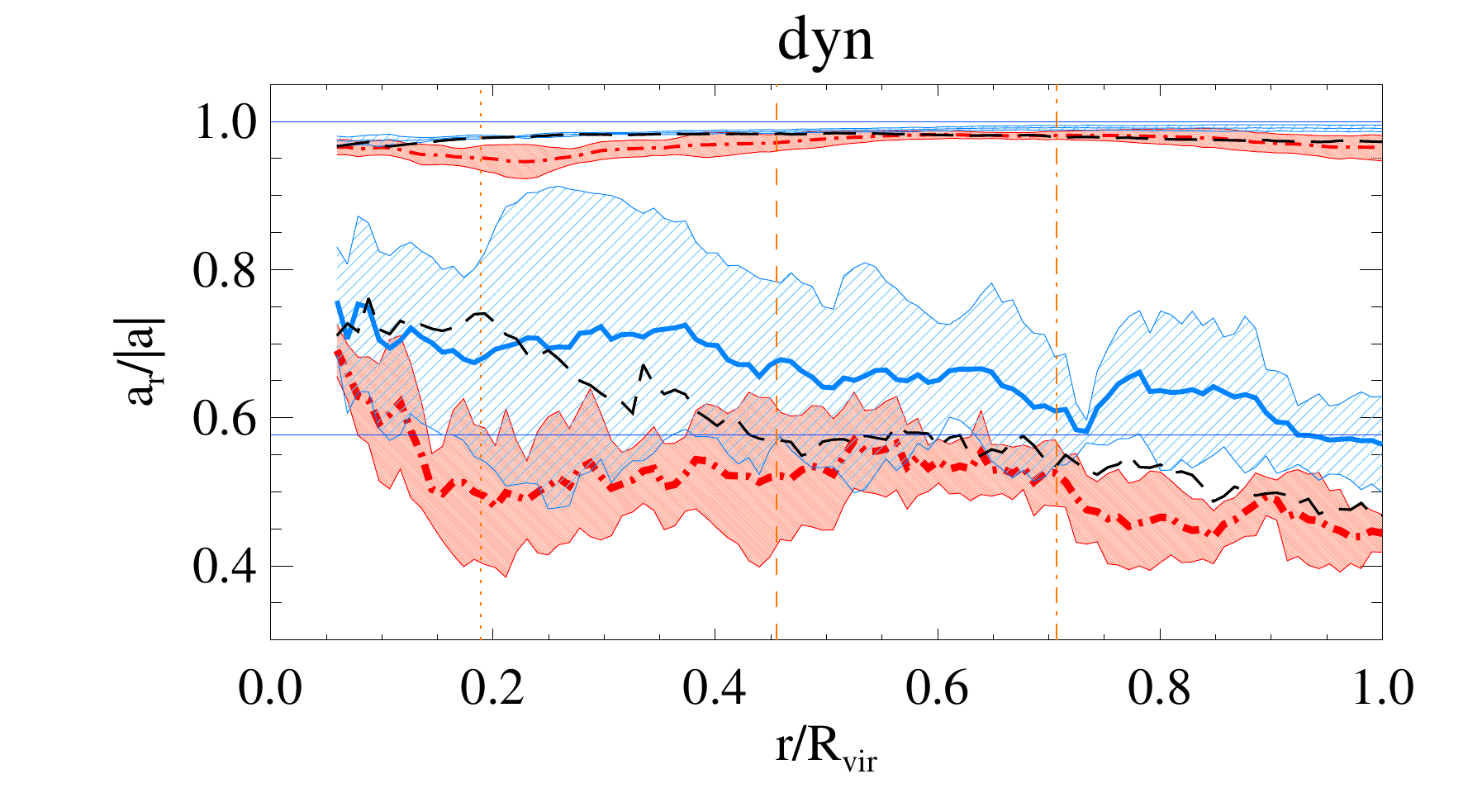}
\caption{Median radial profile of the ratio between the radial
  component of the acceleration and the modulus of the total
  acceleration vector. This ratio quantifies the anisotropy of the ---
  gravitational and hydrodynamical (thin and thick lines,
  respectively) --- acceleration field: $a_r/|{\bf a}| = 1
  \rightarrow$ purely radial, $a_r/|{\bf a}| = 1/\sqrt{3} \rightarrow$
  isotropic (both marked by blue solid lines), and $a_r/|{\bf a}| = 0
  \rightarrow$ purely tangential. Shaded areas represent
  the median absolute deviation from the median value
  in each radial bin.  {\it Upper
    panel:} CC/NCC (blue solid/red dot-dashed line); {\it lower
    panel:} regular/disturbed (blue solid/red dot-dashed line)
  clusters; intermediate systems are marked by the thin black line
  and, for simplicity, no dispersion is marked. From left to right,
  vertical lines mark median values of $\rtwofive$, $\rfive$ and
  $\rtwo$, respectively.%
  \label{fig:acc_anisotr_class}}
\end{figure}
  In fact, we see from Figure~\ref{fig:acc_anisotr_class}~(bottom
  panel) that the gravitational acceleration is generally almost
  radial from the center out to the outskirts (typically \gacc$/|{\bf
    a}_{g}|\sim 0.9$), while the hydrodynamical acceleration shows a
  radial component which decreases with radius (from $\sim0.7$ to
  $\sim0.5$, going from the center to $\rtwo$) and is almost isotropic
  in the intermediate region comprised between $\rfive$ and $\rtwo$.
  This is more evident for disturbed clusters, for which the profile
  of \hacc$/|{\bf a}_{h}|$ is systematically lower, namely less
  radial, than for regular systems.  This off-set is mirrored by the
  one in the $\ratio$ profiles.

  Instead, the same is not observed when the dinstiction between CC and NCC
  systems is adopted, as shown in the upper panel of
  Figure~\ref{fig:acc_anisotr_class}. In this case, the profiles of
  the two populations behave in a very similar way, for both the
  acceleration~components.

\begin{figure}[tb]
\centering
\includegraphics[width=0.49\textwidth,trim=25 0 0 0,clip]{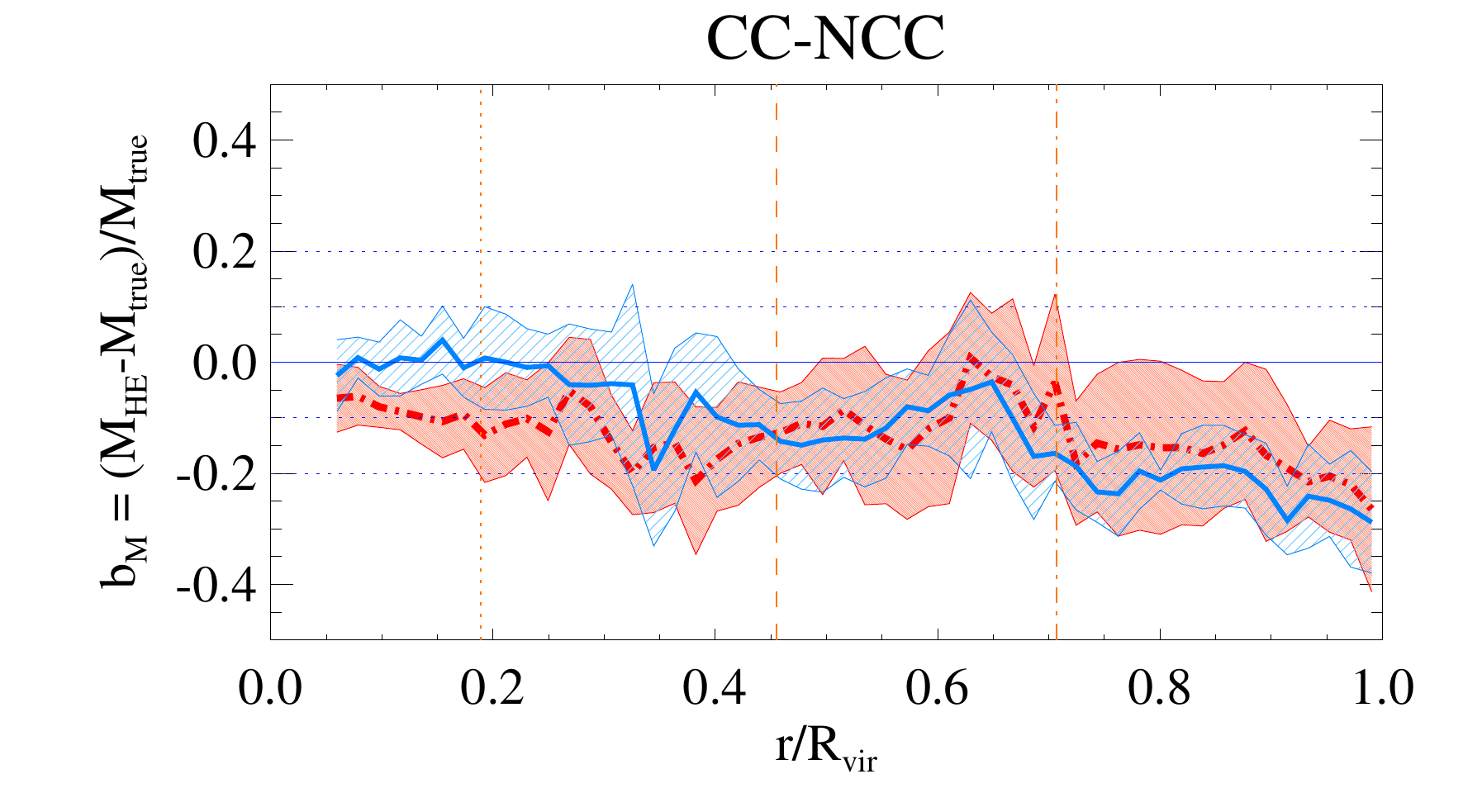}
\includegraphics[width=0.49\textwidth,trim=25 0 0 0,clip]{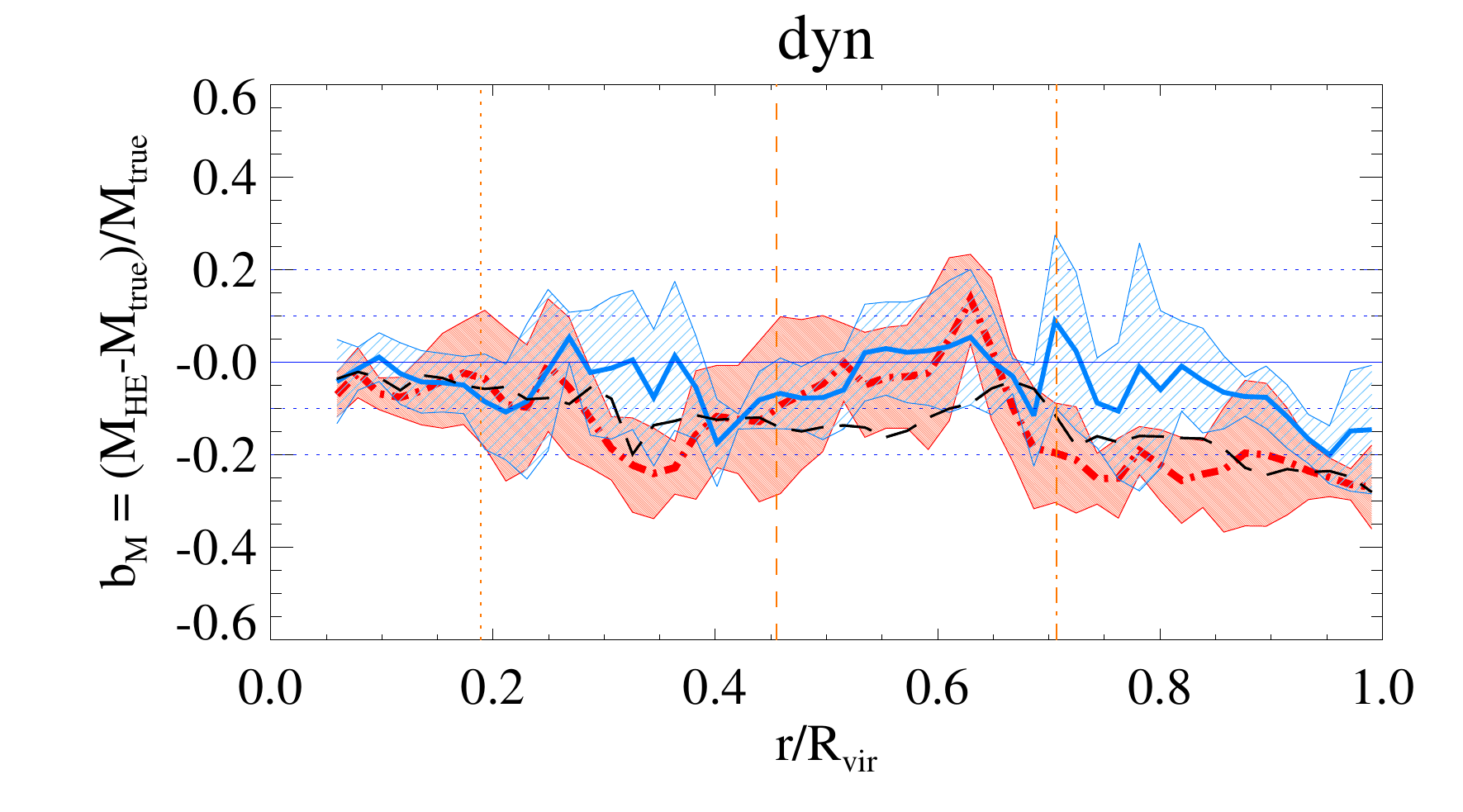}
\caption{Median radial profile of the mass bias, as in
  Figure~\ref{fig:mbias-med}, distinguishing among different cluster
  populations.  {\it Upper panel:} CC/NCC (blue solid/red dot-dashed line); {\it lower
    panel:} regular/disturbed (blue solid/red dot-dashed line)
  clusters; intermediate systems are marked by the thin black line
  and, for simplicity, no dispersion is marked.  The hydrostatic mass,
  $\mhe$, is calculated using $\tmw$. Shaded areas represent
  the median absolute deviation from the median value,
  in each radial bin. From left to right, vertical lines mark median values of
  $\rtwofive$, $\rfive$ and $\rtwo$,
  respectively.\label{fig:mbias-med-class}}
\end{figure}

\enlargethispage*{\baselineskip}
\subsubsection{Hydrostatic mass bias}
Using the same selection criteria to investigate the mass bias we
obtain the results presented in Figure~\ref{fig:mbias-med-class}
(upper and lower panel, respectively).  Here we only show the results
for $\mhe$ computed using the mass-weighted temperature, altough we
verified that using the spectroscopic-like estimate we obtain very
similar profiles, with the only difference of an overall more
significant bias (as seen from Figure~\ref{fig:mbias-med}) and a
larger scatter, especially outside $\rtwo$.

\looseness=-1 We note that the hydrostatic mass bias behaves differently from the
acceleration term with respect to the classification adopted: no
sistematic distinction between regular and disturbed clusters is
evident, except for the outermost region ($>0.7\rvir$).  Instead, a
separation, albeit relatively mild, is found between CC and NCC out to
$\sim\rtwofive$, where there is an offset between their median
profiles and the shaded areas marking the dispersion around the median
values barely touch each other.
In that inner region of the radial profile, the CC population presents
almost zero mass bias while the NCC subsample is characterised by a
mass bias of roughly $10$--$15$~per~cent.
This is mainly due to the different thermo-dynamical properties of the
two classes in the innermost region, where CC clusters are typically
characterized by a higher thermal pressure support with respect to NCC
systems (see Planelles et al., in prep.), despite the similar shape of
their potential well. This then reflects in a better match between the
hydrostatic mass and total gravitating mass.

Interestingly, the comparison between the lower panels of
Figures~\ref{fig:mbias-med-class} and~\ref{fig:acc-med-class}
indicates that the hydrostatic mass bias of disturbed systems is on
average $\lesssim 25$~per cent (with peaks around $25$--$30$~per cent)
despite the larger deviation from $-1$ of $\ratio$ (mostly $\dHE
>20$~per cent, up to $50$~per cent).
The origin of a deviation from zero acceleration (on
  the radial direction) that is larger than the violation of the
  balance between gravitational and thermal pressure forces, must be
  related to gas non-thermalized motions, that are not accounted for
  in our computation of $\nabla P$ (where $P=P_{\rm th}\propto
  \rho\,T$).

\begin{figure}[tb]
\centering
\includegraphics[width=0.49\textwidth,trim=25 0 0 0,clip]{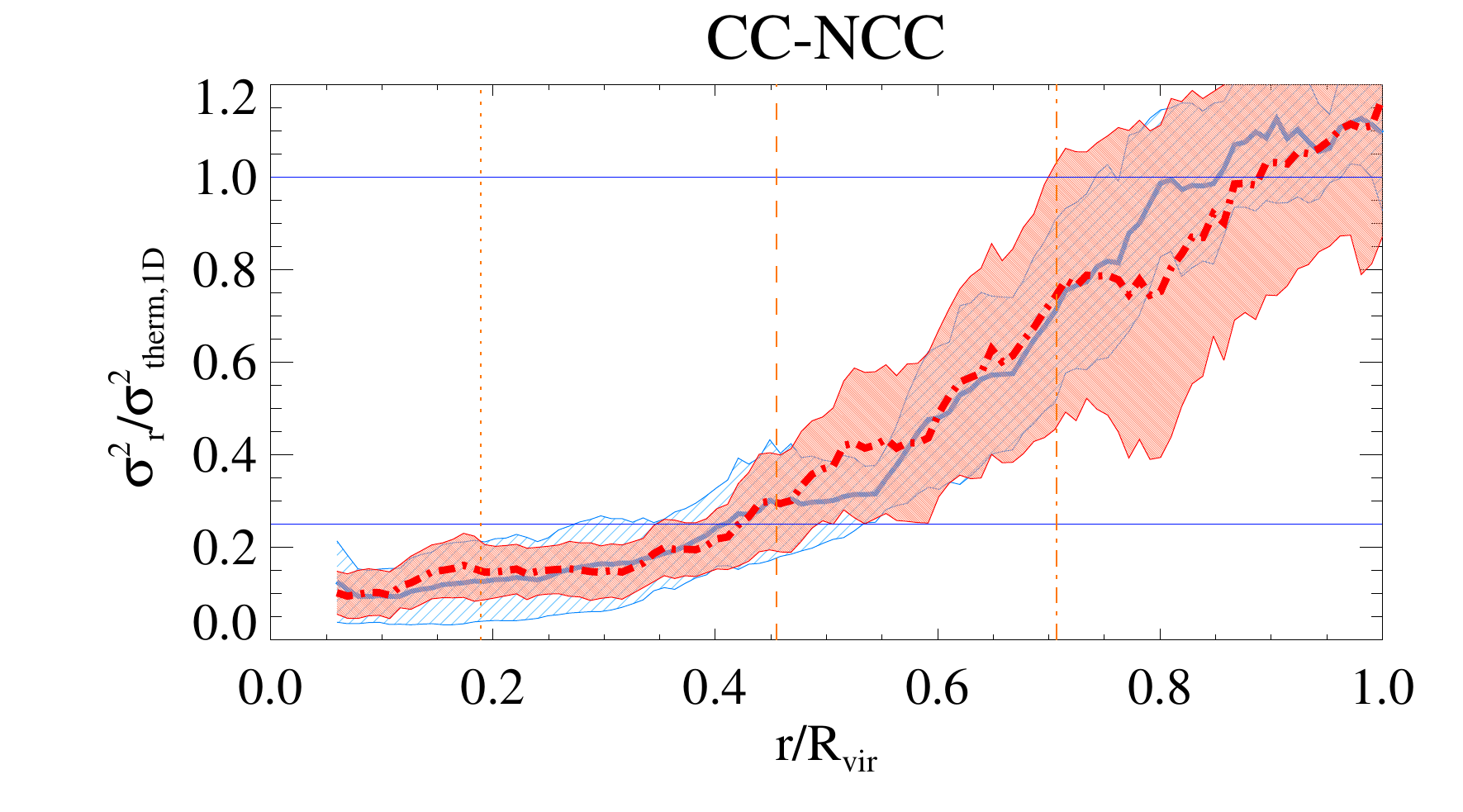}
\includegraphics[width=0.49\textwidth,trim=25 0 0 0,clip]{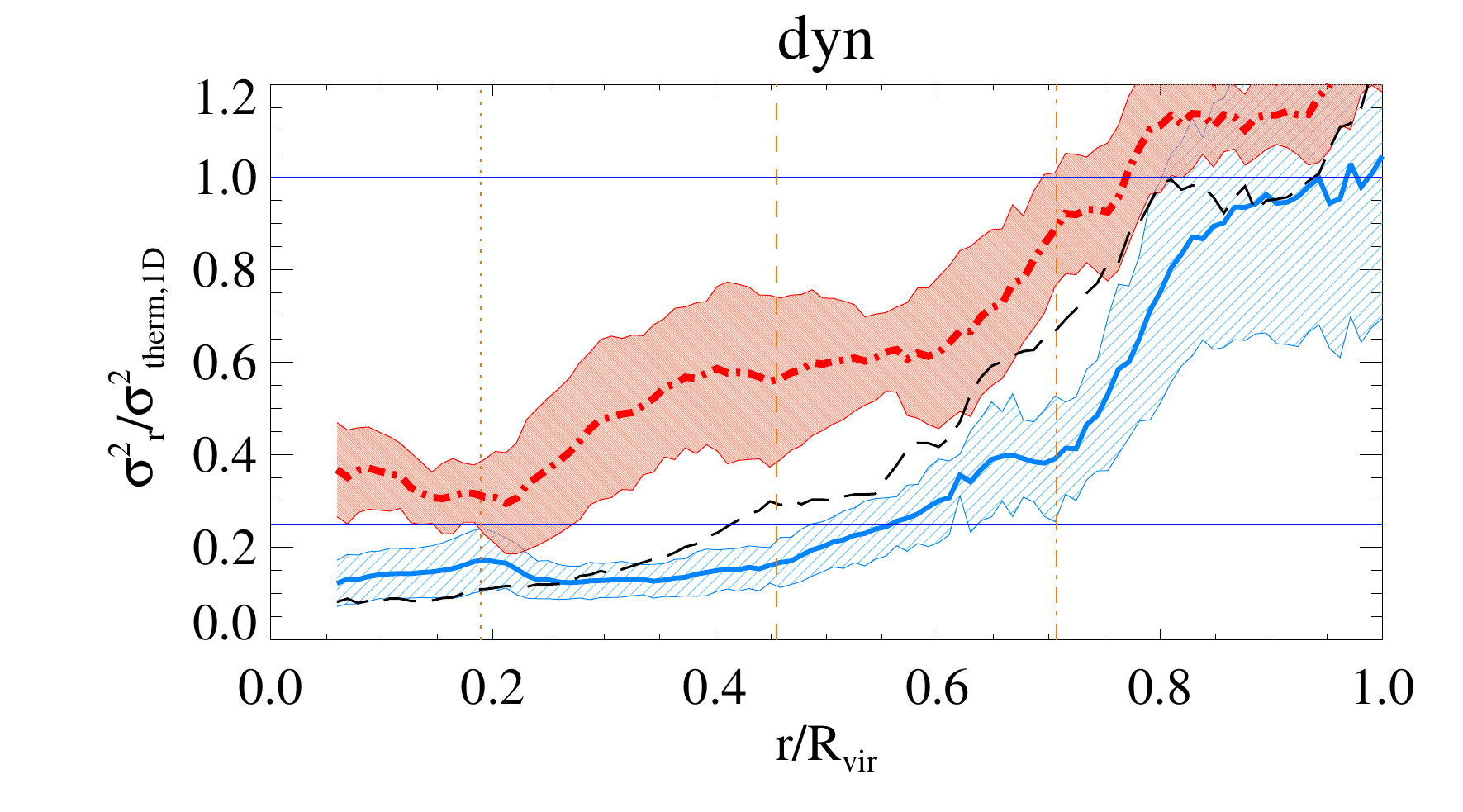}
\caption{Median radial profile of $\sigma^2_r/\sigma^2_{\rm
    therm,1D}$, distinguishing among different cluster populations.
  {\it Upper panel:} CC/NCC (blue solid/red dot-dashed line); {\it
    lower panel:} regular/disturbed (blue solid/red dot-dashed line)
  clusters; intermediate systems are marked by the thin black line
  and, for simplicity, no dispersion is marked. Shaded areas represent
  the median absolute deviation from the median value, in each radial
  bin.  From left to right, vertical lines mark median values of
  $\rtwofive$, $\rfive$ and $\rtwo$,
  respectively.\label{fig:sigr-med-class}}
\end{figure}

From Figures~\ref{fig:acc-med-class} and~\ref{fig:acc_anisotr_class}
(and Figure~\ref{fig:sigr-med-class} below)
we conclude that the radial properties of the ICM acceleration field, and
thus the level of HE, are not very sensitive to the
cool-coreness of the system, but rather depend on its global dynamical
state, whereas the mass bias is more closely related to the cool-coreness,
and so to thermal properties, especially in the central regions
(see Figure~\ref{fig:mbias-med-class}).

\begin{figure*}[tb]
\centering
\includegraphics[width=0.8\textwidth,trim=15 0 0 115,clip]{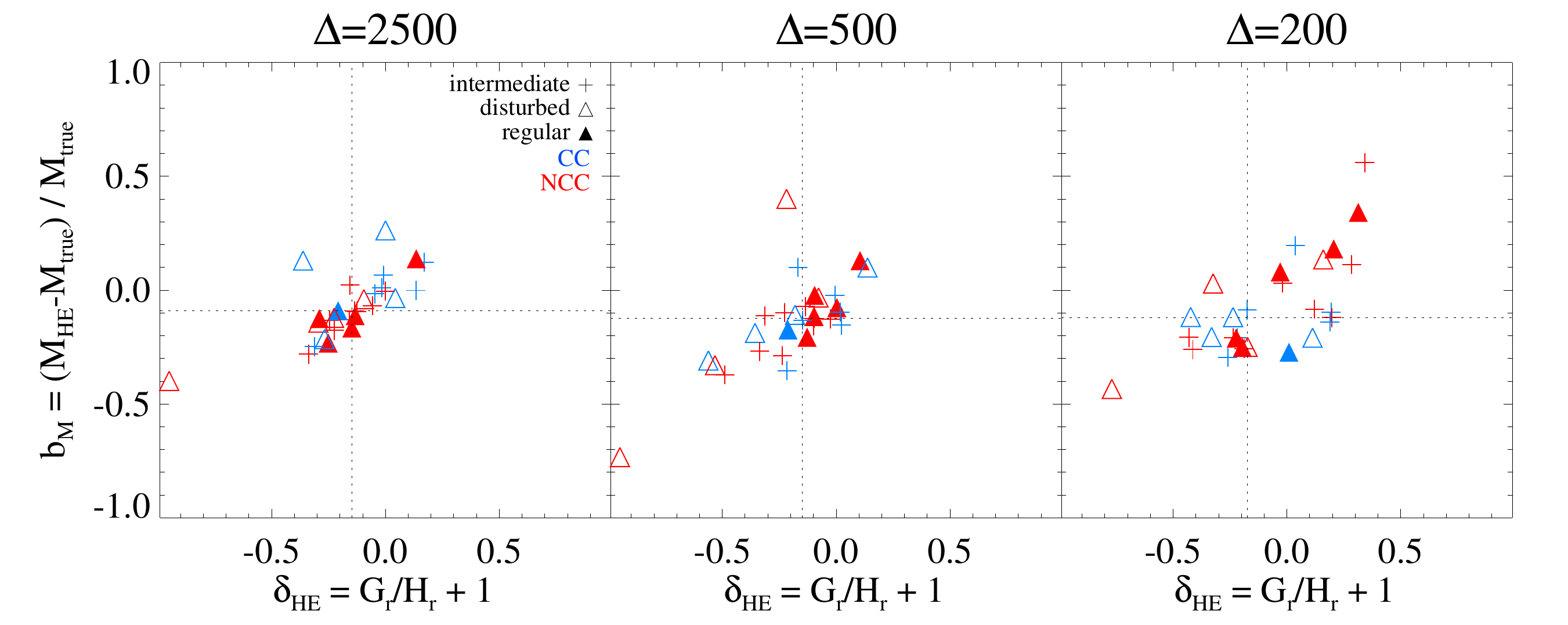}
\caption{Relation between mass bias ($b_M$) and deviation from HE
  ($\dHE$), calculated at $\rtwofive$ (left), $\rfive$ (middle) and
  $\rtwo$ (right).  In each panel median values of the distributions
  are indicated by the dotted lines.\label{fig:mbias-dHE}}
\end{figure*}
\begin{table*}
\centering
\renewcommand{\tabcolsep}{30pt}
\renewcommand\arraystretch{1.1}
\begin{tabular}{|l|c|c|c|}
\hline
& $\Delta=2500$ & $\Delta=500$ & $\Delta=200$\\
\hline
\multicolumn{4}{|c|}{$b_M$}\\
\hline
 all       & $  -0.091 \pm 0.016$ & $ -0.123 \pm 0.015$ & $ -0.120 \pm 0.024$  \\
 CC        & $  -0.001 \pm 0.027$ & $ -0.132 \pm 0.017$ & $ -0.140 \pm 0.020$  \\
 NCC       & $  -0.123 \pm 0.011$ & $ -0.119 \pm 0.021$ & $ -0.102 \pm 0.036$  \\
 regular   & $  -0.118 \pm 0.015$ & $ -0.095 \pm 0.031$ & $ -0.065 \pm 0.080$  \\
 disturbed & $  -0.080 \pm 0.035$ & $ -0.148 \pm 0.060$ & $ -0.162 \pm 0.024$  \\
\hline
\multicolumn{4}{|c|}{$\dHE$}\\
\hline
 all       & $ -0.148 \pm 0.022$  &  $ -0.148 \pm 0.016$  & $ -0.173 \pm 0.039$  \\
 CC        & $ -0.017 \pm 0.046$  &  $ -0.168 \pm 0.049$  & $ -0.175 \pm 0.055$  \\
 NCC       & $ -0.157 \pm 0.019$  &  $ -0.132 \pm 0.024$  & $ -0.101 \pm 0.057$  \\
 regular   & $ -0.178 \pm 0.024$  &  $ -0.096 \pm 0.027$  & $ -0.010 \pm 0.081$  \\
 disturbed & $ -0.247 \pm 0.047$  &  $ -0.287 \pm 0.080$  & $ -0.281 \pm 0.044$  \\
\hline
\end{tabular}
\caption{Median values and errors of the mass bias (top,
  $b_M$) and deviation from HE (bottom, $\dHE$) reported in
  Fig.~\ref{fig:mbias-dHE}, calulated at $R_{2500}$, $R_{500}$
  and $R_{200}$ and for the various subsamples considered.\label{tab:med-val}}
\end{table*}

\bigskip

Differences between the $\ratio$ and mass bias radial profiles can
also be related to the presence of non-thermal, bulk and random,
motions in the gas, as discussed in Section~\ref{subsec:sigr}.  Here,
we present median stacked profiles of $\sigma^2_r/\sigma^2_{\rm
  therm,1D}$ for the subsamples defined on the basis of the cluster
cool-coreness or dynamical classification, in analogy to
Figures~\ref{fig:acc-med-class} and~\ref{fig:mbias-med-class}. From
Figure~\ref{fig:sigr-med-class}, we infer that CC and NCC (upper
panel) behave in a very similar way, with a similar amount of
non-thermal motions increasing towards larger distances from the
center. On the contrary, disturbed systems clearly differentiate from
dynamically regular ones (lower panel) for the presence of a more
substantial amount of radial non-thermal motions with respect to
thermal ones already in the innermost region and out to the virial
radius (systematically higher values of $\sigma^2_r/\sigma^2_{\rm
  therm,1D}$).  So the conclusion is that mass bias and HE-violations
are two different things, the first more related to cool-coreness, the
second more related to the dynamical state of the cluster.

In addition to their radial dependence, it is useful to evaluate
the relation between mass bias and deviation from HE
at interesting distances from the cluster center, such as $\rtwofive$,
$\rfive$ and $\rtwo$ (see Fig.~\ref{fig:mbias-dHE}).  Despite a larger
scatter in the outskirts, the two quantities closely trace each other,
as indicated by the Pearson correlation coefficient for the three
relations: 0.73, 0.72, 0.69, for $\rtwofive$, $\rfive$ and $\rtwo$,
respectively.
The significance of this result is confirmed by corresponding
p-values of the correlation coefficients of $8.1\cdot 10^{-6}$,
$1.1\cdot 10^{-5}$ and $3.3\cdot 10^{-5}$.
In particular, this result is stronger for the subsample
of regular systems, for which the correlation coefficients range from
0.88 at $\rtwofive$ to 0.86 at $\rtwo$
(with p-values of order of 0.02--0.03).
Then, mass bias and violation
of HE are correlated with each other despite reflecting different
aspects of clusters. The outliers of this correlation tend to be
disturbed clusters
and typically reside in the upper envelope of the
relation (higher $b_M$ and lower $\dHE$ than expected from the linear
correlation).

\enlargethispage*{\baselineskip}
\looseness=-1 In Table~\ref{tab:med-val} we report the median values, with 1-$\sigma$ errors,
of the $b_M$ and $\dHE$ distributions shown in Fig.~\ref{fig:mbias-dHE}.
We note that these results correspond to single radial bins (at $\rtwofive$,
$\rfive$ and $\rtwo$, respectively) in the profiles discussed in the
previous sections.

\begin{figure*}[tb]
\centering
\includegraphics[width=0.46\textwidth]{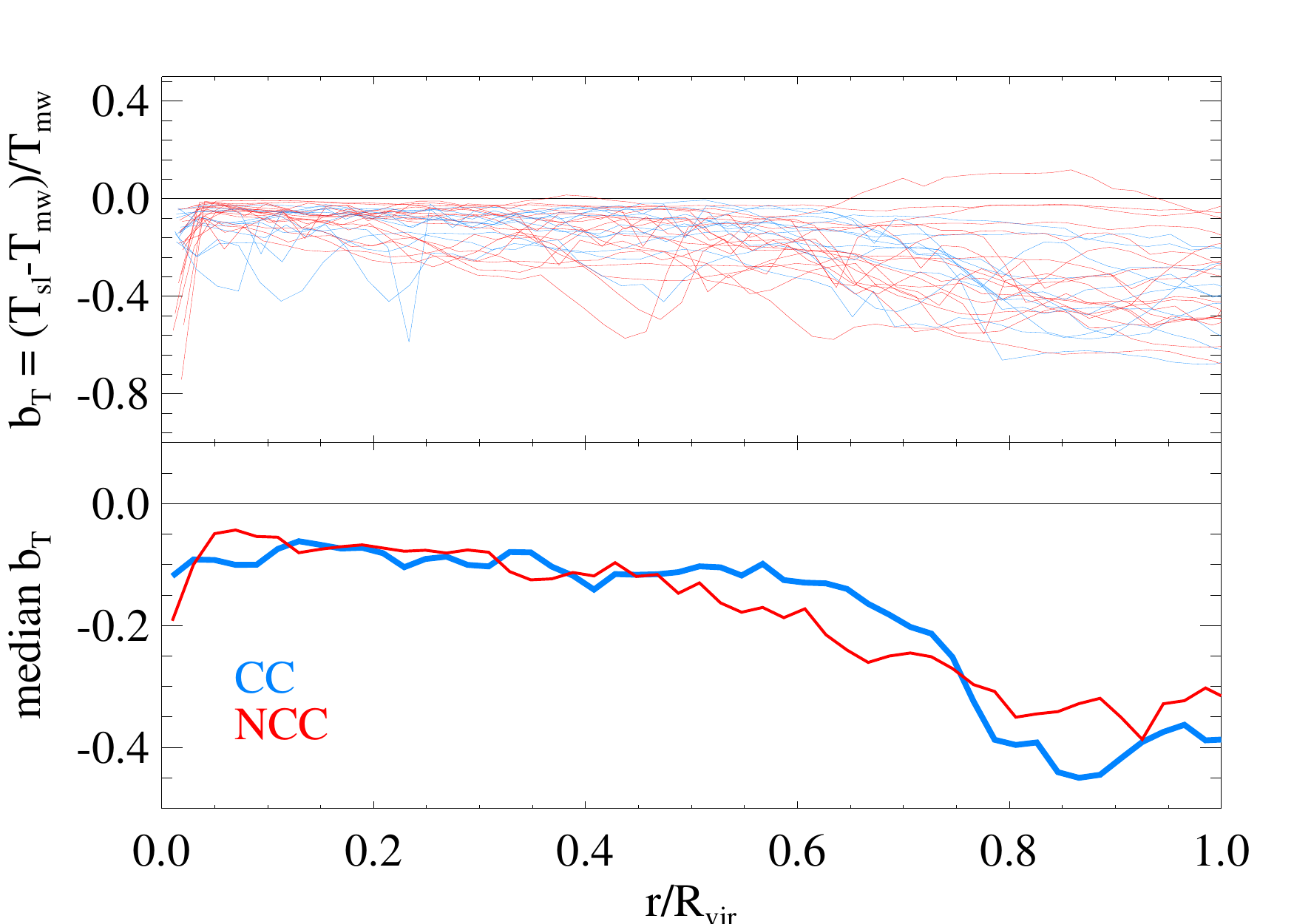}\quad
\includegraphics[width=0.45\textwidth,trim=15 10 10 7,clip]{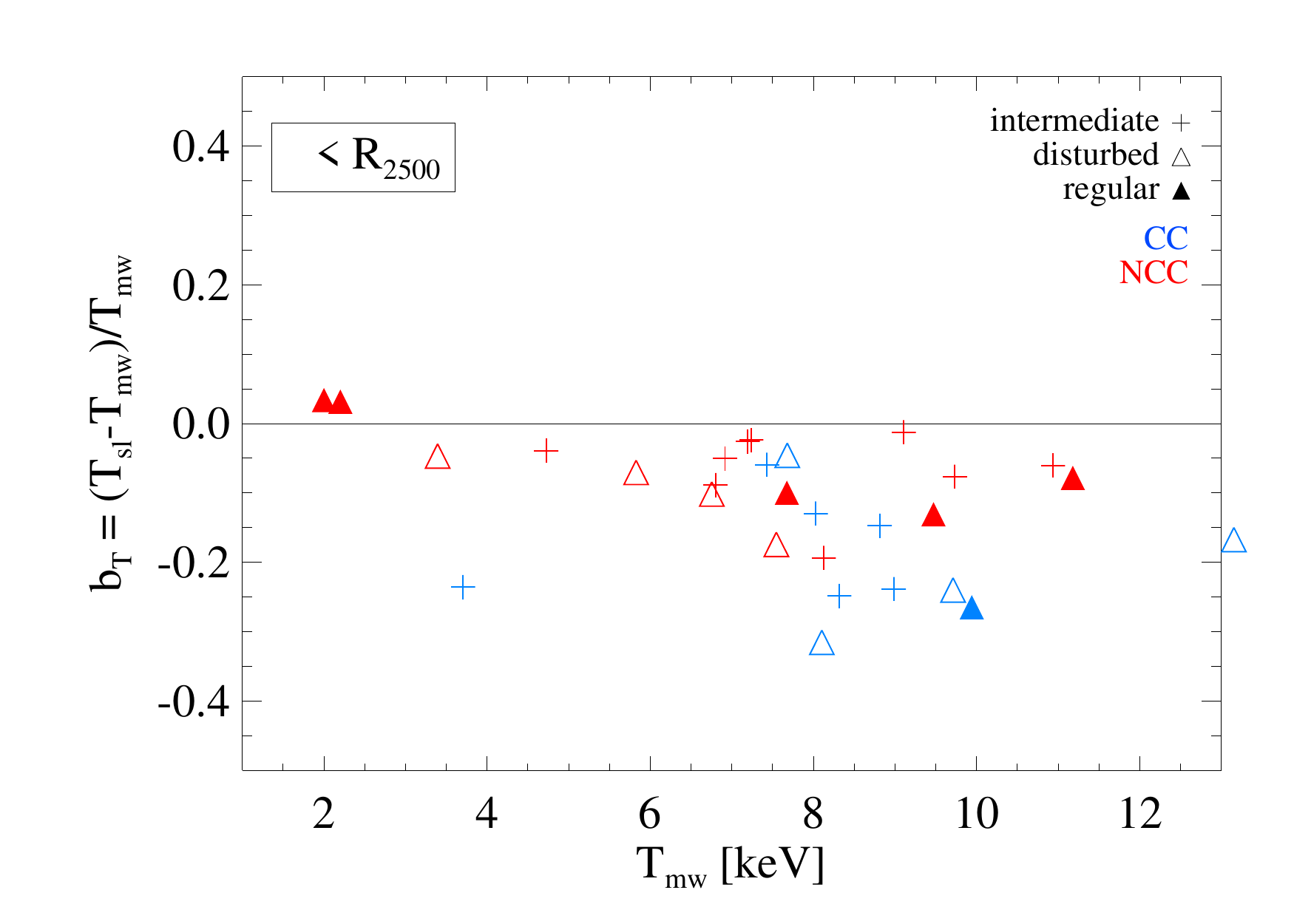}
\caption{%
  {\it Left}: radial profile of the temperature bias $b_T =
  (\tsl-\tmw)/\tmw$, for the 29 main haloes of this work (top
  panel). In the bottom panel the median profiles for CC and NCC are
  shown. Colors refer to CC (blue) and NCC (red) clusters, as in the
  legend. {\it Right}: temperature bias as a function of $\tmw$, for
  the average temperatures within $\rfive$, for the 29 haloes; color
  code marks the CC/NCC (blue/red) classification while symbols
  distinguish among regular, disturbed and intermediate systems.\label{fig:Tbias}}
\includegraphics[width=0.75\textwidth,trim=15 0 0 0,clip]{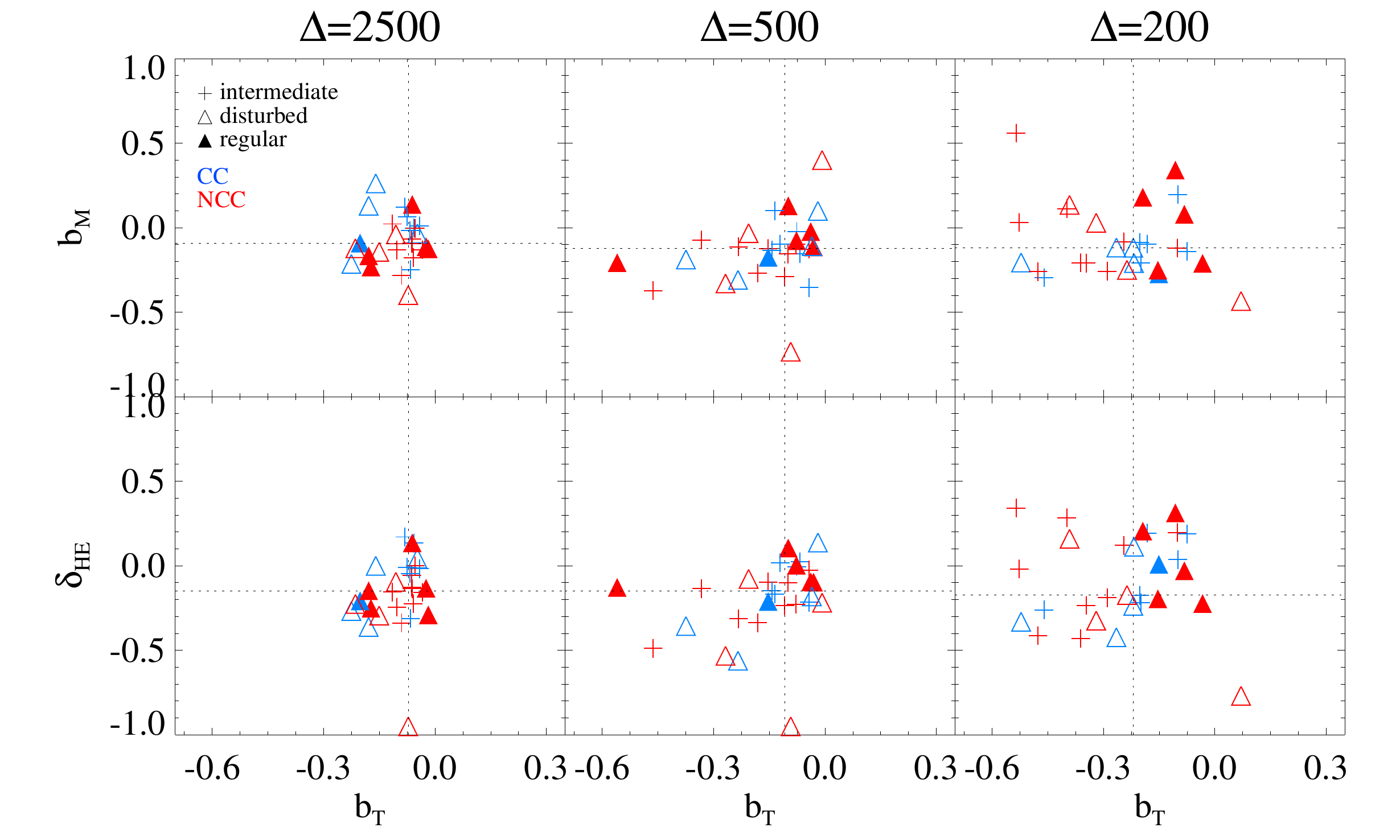}
\caption{Dependence of mass bias ($b_M$; upper panels) and deviation
  from HE ($\dHE$; lower panels) on the temperature bias ($b_T$),
  calculated at $\rtwofive$ (left), $\rfive$ (middle) and $\rtwo$
  (right).  In each panel median values of the distributions are
  indicated by the dotted lines.\label{fig:biases-Tbias}}
\end{figure*}


\subsection{Correlation with temperature bias}

For the purpose of our investigation, it is finally important to
explore the thermal structure of the ICM and the presence of
temperature inhomogeneities, which might affect both the level of
hydrostatic equilibrium and the bias on the hydrostatic mass therefrom
derived.

Numerically, this can be evaluated by comparing the mass-weighted and
spectroscopic-like estimates of temperature, $\tmw$ and $\tsl$, where
the former is a more dynamical measurement while the latter is more
sensitive to the multi-phase nature of the gas.

\looseness=-1 One common way of evaluating this is to calculate the so-called
temperature bias, defined as
\begin{equation}\label{eq:tbias}
b_T = (\tsl-\tmw)/\tmw \,.
\end{equation}
In Figure~\ref{fig:Tbias} we report the radial profile of the
temperature bias for the 29 haloes (left panel, top), for which we
also show the median profiles for the CC and NCC populations,
separately (left panel, bottom).  On average, $b_T$ is always negative out
to the virial radius, indicating that, locally, the spectroscopic-like
temperature typically {\it under}-estimates the dynamical measurement
($\tmw$), at all distances from the cluster center.
Nevertheless, the average bias is found to be quite small,
indicating a relatively homogeneous temperature structure for both
categories, out to $\sim \rfive$ (i.e.\ $\sim 0.4\rvir$), with $b_T \lesssim
10$ per cent.
This can be explained by the improved gas mixing that characterises
these new simulation runs, which allows the gas stripped from the
substructures to efficiently mix and better thermalize with the
surrounding ambient ICM.
In the innermost cluster region the difference between CC and NCC
temperature profiles, decreasing in the first case and flattening or
even rising in the other, is not caught by the temperature bias profile.
The reason for this is that the region sampled by each central radial bin
is not extended enough to capture the central temperature gradient
typical of CC.
In the outer part of the profile, enclosed between $\rfive$ and
$\rtwo$, the bias remains relatively low for CC systems, whereas the
mismatch between $\tsl$ and $\tmw$ increases for NCC.  Towards the
virial radius, the bias significantly increases for both classes.

\enlargethispage*{\baselineskip}
In the right-hand-side panel of Figure~\ref{fig:Tbias} we report
the direct comparison of the global estimates of $\tmw$ and $\tsl$, for
the region enclosed by $\rfive$,
by showing the temperature bias as a function, e.g., of $\tmw$.
Considering the large variety of dynamical states among the clusters
in the sample, we observe on average a very good agreement between the two
values, with $\tsl$ typically underestimating $\tmw$ by only few
percents (within $\rfive$, the median value of the bias is $\sim 5$).
From this relation we also note that there is no evidence for a
dependence of the temperature bias on the global dynamical temperature
of the systems.  In fact, given the well-defined relation between
temperature and mass for the clusters analysed (see Truong et al., in
preparation), we also verified that there is no clear dependence of
the temperature bias on the total cluster mass.


Hydrostatic equilibrium is however a local condition and ultimately
depends on the {\it local} thermodynamical properties of the ICM.
Thus, it is interesting to evaluate the relation between
deviation from HE, mass bias and temperature bias, as in
Figure~\ref{fig:biases-Tbias},
via the dependence of $b_M$ and $\dHE$ (in the upper
  and lower panels, respectively) on the $b_T$, at
  interesting distances from the cluster center, i.e.\
  $\rtwofive$, $\rfive$ and $\rtwo$.

Marking the clusters with different symbols and colors, depending on
their cool-coreness or dynamical classification,
we mainly note a difference between regular and disturbed systems
(filled and empty symbols in the Figure), especially in terms of
scatter, which is significantly larger for the disturbed ones.  This
is particularly evident for the values corresponding to $\rfive$ and
$\rtwo$, where there is a more clear separation between the two
dynamical classes, especially in terms of temperature bias.

Overall, we conclude from Figure~\ref{fig:biases-Tbias} that the local
level of HE and mass bias are not significantly affected by the local
inhomogeneities in the ICM temperature structure.  This is quantified
by very low values of the Pearson correlation coefficients of the
$b_M$-$b_T$ and $\dHE$-$b_T$ relations, which only reach a maximum of
$\sim 0.3$ for $\rfive$ and is very poorly constrained (p-values
$>0.1$).  Only the CC subsample shows evidences for a significant
correlation between $\dHE$ and temperature bias, especially in the
outskirts --- with a Pearson correlation coefficient(p-value) of $\sim
0.68$($0.02$) at $\rfive$ and $\sim 0.66$($0.03$) at $\rtwo$.

\section{Discussion and conclusion}\label{sec:conclusion}

The violation of hydrostatic equilibrium in galaxy clusters has been
widely studied from the numerical point of view, in order to trace
its origin and the connection to the bias in the mass
reconstruction based on the HE hypothesis.

Here, we explored the violation of HE in the ICM by studying the
balance between gravitational and hydrodynamical acceleration, on the
radial direction.  This allowed us to investigate the level of
deviation from HE {\it per se}, i.e. separately from the mass bias, which
additionally implies the assumption of purely thermal pressure
support (with $P_{\rm th}\propto \rho\, T$).

In the following we summarize our main findings.
\begin{itemize}
\item Corrections for HE-violation based on the acceleration term for
  individual clusters are not really achievable. The differences from
  case to case, and depending on the distance from the cluster center,
  make the prediction of a single correction term very challenging,
  even by means of numerical simulations. This is noticeable from the
  significant scatter in the radial profiles of $\ratio$.
\item The classification of relaxed and un-relaxed clusters can be
  misleading, especially when simulations and observations are
  compared: depending on which cluster properties are used to define
  the level of regularity, the differences among the populations range
  from substantial to negligible. Caution is necessary when numerical
  results, e.g.\ scaling relations, are compared to observed ones,
  and vice versa.
\item The acceleration term, quantified via the $\ratio$ ratio, shows
  a systematic difference between the median radial profile of dynamically
  regular clusters and that of disturbed ones, with the latter showing
  a larger deviation from HE ($\dHE >20$~per cent), i.e. from $d{\bf
    v}/dt = 0$ (on the radial direction). This is especially
  significant in the outskirts ($\dHE \sim 50$~per cent).

  Instead, we find no clear dependence of the $\ratio$ profile on the
  system cool-coreness, from comparing CC and NCC median profiles.
\item On the contrary, when the hydrostatic mass bias is concerned,
  CC and NCC clusters behave differently, especially in the
  inner region ($\lesssim\rtwofive$), whereas no siginificant
  distinction is observed between the mass bias of regular and
  disturbed clusters, given the large \mbox{dispersion.}
\item Typically, we find that the clusters for which the radial
  profile of mass bias and deviation from HE ($\dHE$) poorly trace
  each other present a significant amount of non-thermal (bulk and
  random) gas motions with respect to thermal ones, in the radial
  direction, quantified by $\sigma^2_r/\sigma^2_{\rm therm,1D}>0.3$
  already in the innermost regions.
\item We find also a clear correlation between values of the hydrostatic
  mass bias and the deviation from HE computed at $\rtwofive$,
  $\rfive$ and $\rtwo$, with the main outliers in this picture
  represented by dynamically disturbed systems. From this we conclude
  that the local deviation from HE is of order 15--20 per cent
  (increasing towards the outskirts), and it is generally well traced
  by the local mass bias (of order 10--15 per cent).
\item The temperature structure of the clusters in the sample appears
  to be relatively regular, with a temperature bias lower than what
  previously found in SPH simulations. In fact we find that $\tsl$
  typically underestimates $\tmw$ by few percents in the innermost
  region, increasing up to $\sim 15$--20~per cent towards the outskirts.
\item On average, we find no strong correlation
  between the local dishomogeneity in the thermal structure
  (quantified by the temperature bias) and the local deviation from HE
  or mass bias.
\end{itemize}

Simulations are extremely powerful for in-depth studies like the one
presented in our analysis, since the HE validity in the ICM of clusters can be
explored in full detail cluster by cluster.
In particular, we have shown different levels of deviation from HE and
of hydrostatic mass bias for various cluster populations, classified
on the basis of their global dynamical state --- as often done in
simulations --- and core thermal properties --- as typically done in
observations.
This was possible by employing state-of-the-art cosmological
simulations that include the description of several hydrodynamical
processes taking place in galaxy clusters and, most importantly, that
were able to generate the observed co-existence of cool-core and non-cool-core
systems with thermo- and chemo-dynamical properties in good agreement
with observations~\cite[][]{rasia2015}.
We have shown that CC and dynamically-regular clusters are very
different populations in terms of HE-deviation and mass bias, and
similarly NCC clusters clearly differ from disturbed systems in the
same respect.

\looseness=-1 Nevertheless, such numerical studies also remark the intrinsic
difficulty of predicting from simulations an accurate correction to
X-ray based (or more in general to hydrostatic) masses on a
cluster-by-cluster basis. Still, the virtue of simulations is that
they allow us to calibrate such a correction in a statistical sense,
through the calibration of scaling relations between true masses and
hydrostatic masses. Clearly, the reliability of these corrections, in
view of their application to precision cosmology with clusters,
depends on the degree of realism of the simulations.
In this respect, additional forces, not treated in our work, should
also be taken into consideration in cluster simulations, such as
magnetic field and cosmic ray pressure, which alter the momentum of
the intra-cluster gas in real clusters and can contribute to the
support against the gravitational force.

From the observational side, up-coming (ASTRO-H) and future
(e.g.\ Athena) X-ray missions, thanks to their high-resolution
spectrometry capabilities,
will help to better characterize the various terms of pressure
support against gravity in clusters.  This will be achieved via
measurements of the gas velocities from non-thermal broadening and
center-shifts of spectral emission lines from heavy ions, e.g.\ Iron.
Even though the chance to measure gas accelerations still remains
remote, if not impossible, future X-ray observations will likely
permit to reduce and control the effect due to the assumption of HE
and to obtain more accurate mass estimates, especially from
spatially-resolved observations.

Additionally, given the distinct level of deviation from HE depending
on the cluster dynamical state, rather than on their cool-coreness, the
possibility to observationally constrain the amount of non-thermal
motions in the ICM could provide a new, complementary way of
classifying cluster populations.

\acknowledgments
We are greatly indebted to Volker Springel for giving us access to the
developer version of the GADGET3 code. We acknowledge financial
support from PIIF-GA- 2013-627474, NSF AST-1210973, PRIN-MIUR
201278X4FL, PRIN-INAF 2012 ``The Universe in a Box: Multi-scale
Simulations of Cosmic Structures'', the INFN INDARK grant, ``Consorzio
per la Fisica'' of Trieste, CONICET and FONCYT, Argentina. Simulations are
carried out using Flux HCP Cluster at the University of Michigan,
Galileo at CINECA (Italy), with CPU time assigned through ISCRA
proposals and an agreement with the University of Trieste, and PICO at
CINECA though our expression of interest.
S.P. acknowledges support by the Spanish Ministerio de Economia y
Competitividad (MINECO, grants AYA2013-48226-C3-2-P) and the
Generalitat Valenciana (grant GVACOMP2015-227).
M.G. is supported by NASA through Einstein Postdoctoral Fellowship
Award Number PF-160137 issued by the Chandra X-ray Observatory Center,
which is operated by the SAO for and on behalf of NASA under contract
NAS8-03060.
A.M.B. is supported by the DFG Research Unit 1254 'Magnetisation of
interstellar and intergalactic media' and by the DFG Cluster of
Excellence 'Universe'.
Also, we would like to thank the anonymous referee for a careful
reading of the manuscript and constructive comments that helped
improving the presentation of this work.
%

\appendix

\section{The hydrostatic sphere test}\label{app:hyd_sph}

In order to test the approach used in this work, we set up a
sphere in hydrostatic equilibrium as a study case and apply the analysis
of the acceleration components presented above.

The hydrostatic sphere is set up with a total virial mass of $4\cdot
10^{14}\msunh$, resolved with 369300 DM particles and 403508 gas
particles. The mass resolution is
$m_{\rm DM}=3.1\cdot 10^9\msun$ and $m_{\rm gas}=6.2\cdot 10^8\msun$,
for DM and gas respectively.  The simulation of the hydrostatic sphere
has been performed with the same version of the code used for the
other haloes analysed in this work, but including only non-radiative
hydrodynamics. It has been evolved for a large enough number of dynamical time-steps till an ideal configuration of hydrostatic equilibrium is reached.
\begin{figure}[tb]
\centering
\includegraphics[width=0.45\textwidth]{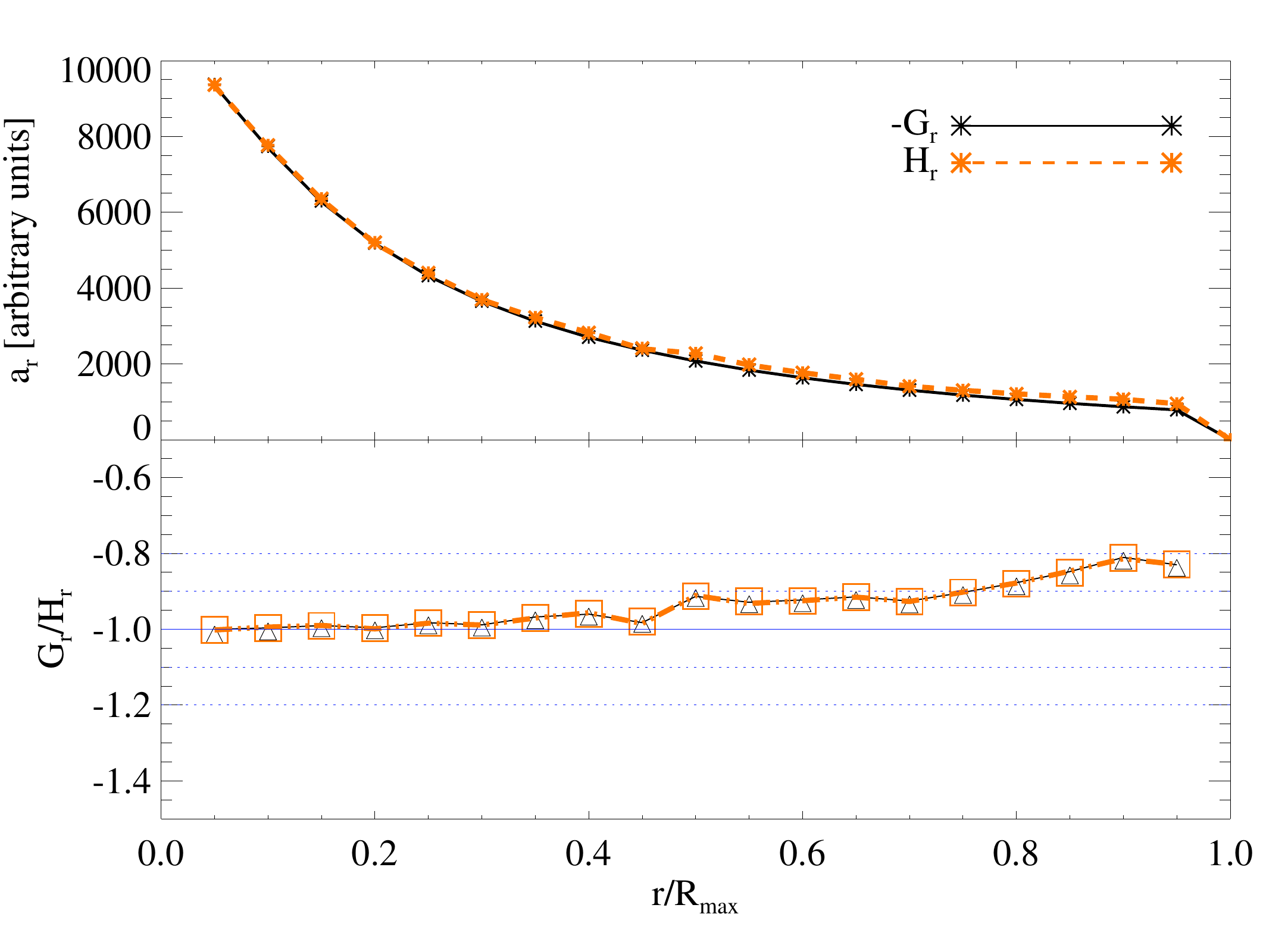}
\caption{Hydrostatic sphere. {\it Top panel:} radial profiles of the \gacc\ (changed in sign, for visualization and comparison purposes) and \hacc\ components, as in the legend. {\it Bottom panel:} radial profile of the (mean and median) $\ratio$ profile, both from the particle-based calculation (black trinagles and solid line) and from the two separate components profiles (orange squares and dot-dashed line). In both panels, symbols and lines indicate mean and median values, respectively.\label{fig:hyd_sphere}}
\end{figure}

As visible from the \gacc~(changed in sign) and \hacc\ profiles in
Figure~\ref{fig:hyd_sphere}\,(upper panel), the HE configuration shows
very good balance between the two components out to very large radii,
with their ratio showing almost perfect balance in the central region
($\dHE \lesssim 5\%$), and presenting deviations from HE smaller than
10\% roughly out to $\rtwo$ ($\sim 1200\kpc$, i.e. $0.8\, R_{\rm max}$ in
the Figure).
  In fact, from the inspection of the $\ratio$ profile at different
  time-steps, till the final configuration shown in
  Figure~\ref{fig:hyd_sphere}\,(lower panel), we observe a clear trend
  of the profile to set towards the -1 reference line, with the radial range in
  which the equilibrium condition is satisfied extending outwards.
  Given this, we expect that a perfect HE profile would be reached
  after an ideally large number of dynamical times.
  From the comparison with the various profiles obtained for the
  cosmological cases, presented in the previous sections, we estimate
  this effect not to cause any bias to the conclusions drawn from our
  study.

  This study case is used to confirm that the HE, on the radial
  direction, corresponds indeed to $\ratio \sim -1$, i.e. to the balance
  of the radial components of gravitational and hydrodynamical
  accelerations.

From the comparison in the Figure between the profiles
of \gacc, \hacc\ and $\ratio$, we additionally tested that:
\begin{enumerate}[(i)]
\item the particle-based approach and the use of the separate profiles
  of \gacc\ and \hacc\ converge to the same result for an ideal
  hydrostatic gas distribution (perfect overlap between symbols (mean values) and lines (median values) in the lower panel);
\item mean and median values within the radial bins provide exactly
  the same result (perfect overlap between curves and symbols), given
  the absence of gas inhomogeneities.
\end{enumerate}

%
\bibliographystyle{apj}
\bibliography{bibl}


\end{document}